\documentclass[sigplan,authorversion,nonacm]{acmart}
\usepackage{multirow}
\usepackage{subcaption}

\usepackage{amssymb}  

\usepackage{enumitem} 
\usepackage{tikz}
\usepackage{xcolor}
\newcommand*\colorcircled[2]{%
  \tikz[baseline=(char.base)]{
    \node[shape=circle, fill=#1, minimum size=11pt, inner sep=0pt] (char)
      {\resizebox{!}{5pt}{\textcolor{black}{#2}}};}}

\newcommand*\orangecircled[1]{\colorcircled{yellow!55!orange}{#1}}

\AtBeginDocument{%
  }

\begin{document}

\title{Vmem: A Lightweight Hot-Upgradable Memory Management for In-production Cloud Environment}

\author{Hao Zheng\textsuperscript{1*},\quad
Qiang Wang\textsuperscript{2*},\quad
Longxiang Wang\textsuperscript{2},\quad
Xishi Qiu\textsuperscript{1},\quad
Yibin Shen\textsuperscript{1},\quad
Xiaoshe Dong\textsuperscript{2},\quad
Naixuan Guan\textsuperscript{1},\quad
Jia Wei\textsuperscript{2},\quad
Fudong Qiu\textsuperscript{1},\quad
Xingjun Zhang\textsuperscript{2},\quad
Yun Xu\textsuperscript{1},\quad
Mao Zhao\textsuperscript{2},\quad
Yisheng Xie\textsuperscript{1},\quad
Shenglong Zhao\textsuperscript{1},\quad
Min He\textsuperscript{1},\quad
Yu Li\textsuperscript{1},\quad
Xiao Zheng\textsuperscript{1},\quad
Ben Luo\textsuperscript{1},\quad
Jiesheng Wu\textsuperscript{1}}
\thanks{\textsuperscript{*} Hao Zheng and Qiang Wang contributed equally to this work.}

\affiliation{
\textsuperscript{1} Alibaba Cloud, Hangzhou, \country{China} \\
\textsuperscript{2} School of Computer Science, Xi'an Jiaotong University, Xi'an, \country{China}
}

\begin{abstract}
Traditional memory management suffers from metadata overhead, architectural complexity, and stability degradation, problems intensified in cloud environments. Existing software/hardware optimizations are insufficient for cloud computing’s dual demands of flexibility and low overhead. This paper presents Vmem, a memory management architecture for in-production cloud environments that enables flexible, efficient cloud server memory utilization through lightweight reserved memory management. Vmem is the first such architecture to support online upgrades, meeting cloud requirements for high stability and rapid iterative evolution. Experiments show Vmem increases sellable memory rate by about 2\%, delivers extreme elasticity and performance, achieves over 3$\times$ faster boot time for VFIO-based virtual machines (VMs), and improves network performance by about 10\% for DPU-accelerated VMs. Vmem has been deployed at large scale for seven years, demonstrating efficiency and stability on over 300{,}000 cloud servers supporting hundreds of millions of VMs.
\end{abstract}



\maketitle
\pagestyle{plain}

\section{Introduction}
Cloud computing underpins modern IT infrastructure, improving resource utilization and flexibility via on-demand resources. Virtualization \cite{jain2016overview,jia2023making}, its core, abstracts physical resources into multiple VMs, enabling efficient use and isolation in large-scale, multitenant settings with dynamic allocation. Memory virtualization, a critical component, lets multiple VMs efficiently share and manage physical memory. Thus, effective, flexible, and reliable memory virtualization is essential in cloud environments. However, current technologies face the following challenges:

\textbf{Lightweight and Hot Upgrading}: 
In mainstream memory virtualization, the hypervisor allocates VM memory via OS interfaces. OS memory management handles partitioning, paging, dynamic (de)allocation, isolation, protection, and caching to ensure efficient, stable operation across diverse platforms and workloads~\cite{gorman2004understanding,huang2016evolutionary}. Virtualization’s reliance on these complex, general-purpose schemes adds significant implementation and maintenance overhead, while many features are unnecessary in clouds and may reduce server stability. This complexity tightly couples memory management to the OS, lacking online hot-upgrade capability and failing to meet requirements for rapid fixes, user transparency, and iterative memory usage in cloud. For cloud providers, hot-upgradability is essential for handling failures and updates~\cite{zhang2019fast}. Thus, a lightweight, hot-upgradable memory management design tailored to cloud computing is needed to replace traditional heavyweight approaches.

\textbf{Sellable Memory Rate}:
Traditional memory management uses segmented paging, incurring substantial metadata to track transparent huge pages, slub allocators, and file mappings. In cloud environments, much VM memory metadata is redundant for the host OS. Growing memory demand from modern AI workloads forces providers to increase host memory, sharply raising costs. Server memory now exceeds TB level, with metadata overhead alone around 20\,GB, reducing VM-available memory and sellable rates. Mixed usage between the host OS and VMs leads to fragmentation \cite{kim2015controlling} and leaks \cite{jung2014automated}, further reducing VM-available memory. To ensure stability, large memory is reserved for the host OS even if unused. While reserved memory can reduce metadata overhead \cite{dmemfs}, host OS and reserved memory remain isolated, preventing sharing and making it hard to shrink host OS memory to increase sellable memory. A new memory management method is needed to increase sellable memory on cloud servers without affecting stability.

\textbf{Extreme Performance and Rapid Elasticity}:
To optimize virtualization performance, huge pages are widely adopted~\cite{jia2023making}, but fragmentation reduces availability over time~\cite{panwar2018making}. Conventional methods reserve separate memory pools for different granularities at boot, leading to unstable maximum allocations, blocking dynamic pool sharing, and preventing large granularities from matching sellable VM specifications—making them unsuitable for diverse, elastic cloud workloads.  
With widespread device passthrough for near bare-metal I/O~\cite{amit2015bare,dong2012high}, all VM's memory must be preallocated and mapped for DMA. Existing huge-page requires extensive allocation and mapping at VM boot with passthrough devices, significantly delaying large-VM creation and impairing rapid scaling and user experience. For data security, all VM memory must be zeroed, and traditional boot-time zeroing further adds delay.  
Dynamically allocating huge pages of varying granularities, maximizing performance under extreme conditions, and supporting on-demand allocation with rapid provisioning and reclamation in cloud environments present a challenge for memory management.

Currently, optimization strategies for cloud memory management include reducing metadata overhead~\cite{HVO,dmemfs}, improving allocation algorithms~\cite{kwon2023efficient,zhang2023partial,schrammel2022jenny,xie2022cetis,Tiered-Memory-Management,10125028}, and enhancing address translation efficiency~\cite{zhang2024direct,alverti2020enhancing,margaritov2021ptemagnet,stojkovic2022parallel}. However, these approaches cannot fully address the above problems, and most require kernel code modifications, increasing complexity and failing to meet cloud requirements for lightweight design, high sellable memory rate, extreme performance, and hot-upgradability. We therefore propose Vmem, a lightweight, hot-upgradable memory management architecture for cloud environments that enables efficient resource management and supports hot upgrades via modular design. This paper makes the following contributions:

(1) \textbf{Modular and Hot-upgradable Memory Management Framework}: We propose a novel lightweight framework that uses kernel modularization to directly manage VM memory, eliminating the heavy burden of traditional general-purpose OS. This is the first framework to support online hot-upgrading of memory management, addressing the rapid iteration needs of cloud computing.

(2) \textbf{Inventory-friendly reserved memory management}: We implement reserved memory management that reduces the large metadata overhead of traditional approaches and allows dynamic sharing of reserved memory with the host OS, enabling maximal and deterministic allocation of sellable memory to meet inventory demands and greatly increasing sellable memory for cloud services.

(3)	\textbf{Mixed-Grained Memory Allocation and FastMap Mechanism}: Leveraging lightweight reserved memory slicing, we design a mixed-grained memory allocation and mapping mechanism, coupled with a FastMap bidirectional mapping mechanism. This approach maximizes the performance potential of huge page memory while significantly enhanced the rapid elasticity capabilities of VM.

(4)	\textbf{Large-Scale Production Deployment}: Vmem has been adopted as the default memory management component in our public cloud infrastructure. It has been running on multiple hardware platforms and kernel versions for over seven years, serving more than 300,000 servers and supporting over one billion VMs.

\section{Motivation}

\subsection{Challenges of Traditional Memory Management}

\subsubsection{Stability Issues}
\begin{figure}[htbp] 
   \centering
   \includegraphics[width=0.9\linewidth]{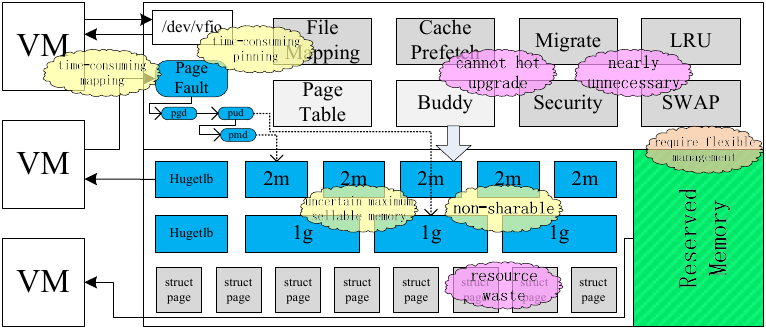}
   \caption{The Limitations of Linux Memory Management.} 
   \label{fig:mm} 
\end{figure}

Traditional Linux memory management targets general-purpose workloads \cite{gorman2004understanding,huang2016evolutionary}, with features such as address space management, page table handling, multi-granularity allocators, page reclamation, swapping, file mapping, prefetching, fault handling, security, and debugging. In cloud environments, most server memory is allocated to VMs, making most of these features unnecessary (gray area in Figure \ref{fig:mm}). Nonetheless, mainstream cloud services continue to rely on  generic Linux memory management, inheriting substantial unused complexity that increases code size, fragility, and stability risks \cite{mi2019skybridge}.

Once deployed, cloud servers running customer workloads cannot be easily taken offline for OS updates, yet critical bugs must still be fixed. Traditional memory management is statically compiled into the kernel and cannot be hot-upgraded in production, allowing only limited hot patches \cite{kpatch14,zhou2020kshot} that are insufficient for robust stability. As shown in Table \ref{tab:linux-mem-bug}, our production cloud environment uses several mainstream Linux LTS kernels, each accumulating numerous memory-related patches over time, including dozens for critical crashes or security vulnerabilities. Relying solely on hot patches incurs high maintenance overhead and cannot resolve all issues, making long-term system stability a major challenge.

\begin{table}[htbp]
\caption{Statistics of Memory Management-related Patches.}
\label{tab:linux-mem-bug}
\scriptsize
\setlength{\tabcolsep}{10pt}
\begin{tabular}{lllll}
\toprule
\textbf{Linux LTS} & \textbf{Commits} & \textbf{Bugfixes} & \textbf{Panics} & \textbf{CVEs} \\
\midrule
4.9   & 592  & 460 &  79 &  82 \\
4.19  & 650  & 487 & 107 &  78 \\
5.10  & 402  & 237 &  37 &  28 \\
\bottomrule
\end{tabular}
\end{table}

\subsubsection{Metadata Overhead}
Traditional operating systems maintain a \texttt{struct page} for each 4KB physical page to support memory management. Designed for diverse general-purpose functions, this structure is highly complex. Even after optimization, each \texttt{struct page} occupies 64 bytes per 4KB page, accounting for \texttt{1.56\%} overhead. In VMs, most fields are unused, yet it remains the dominant source of memory metadata overhead. On a typical early-generation 384\,GB cloud server, this single overhead exceeds 6\,GB - excluding other metadata - and is equivalent to the memory capacity of one or more sellable VMs. This waste grows rapidly with increasing server memory capacity and scale.

\subsection{Challenges in Cloud Memory Management}
\subsubsection{Performance-Benefit Trade-offs}
\label{huge_perf}
\begin{figure}[htbp] 
   \centering
   \includegraphics[width=0.95\linewidth]{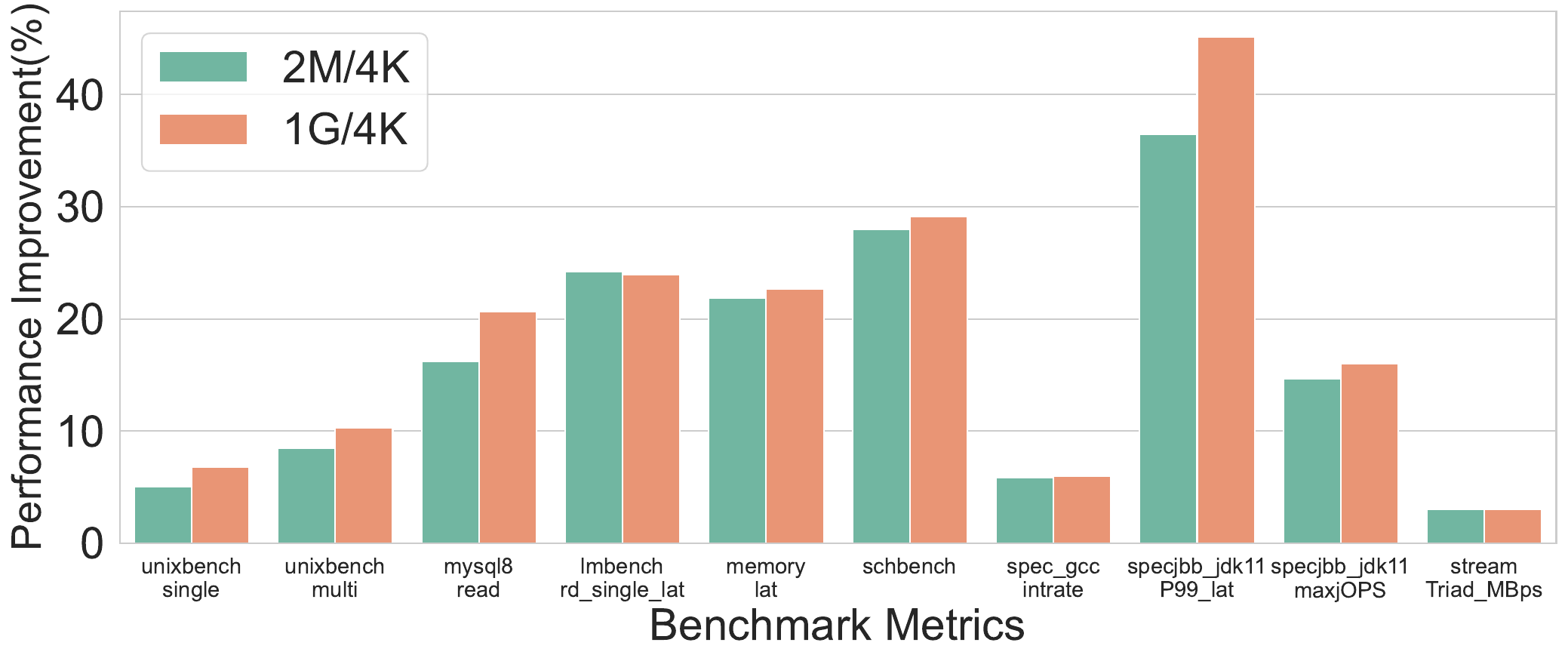}
   \caption{Performance Improvement of Huge Page.} 
   \label{fig:huge} 
\end{figure}

Memory usage in cloud is relatively simple: the hypervisor allocates memory to VMs and sets up page-table mappings. The most widely used feature is Hugetlb \cite{litke2007turning,lu2006using}, as huge pages (e.g., 2MB, 1GB) reduce TLB misses and improve performance over 4KB pages. As shown in Figure \ref{fig:huge}, both 2MB and 1GB pages outperform 4KB pages, but 1GB pages offer only marginal benefit over 2MB—except for a few workloads such as \texttt{SPECjbb}—because TLB capacity is sufficient for 2MB pages in typical cloud servers. Thus, for performance reasons, providers typically use 2MB Hugetlb. However, Hugetlb still incurs substantial metadata overhead, as each huge page consists of small pages, each with a \texttt{struct page} overhead.

Although 1\,GB pages offer little performance gain over 2\,MB pages in most cases, certain performance-critical workloads still require them, especially when TLB entries are limited. However, Hugetlb manages different huge-page sizes separately with fixed page-table mappings, limiting flexibility and preventing resource sharing across page-size pools (blue area in Figure \ref{fig:mm}). For example, if 1\,GB pages are reserved, their resources cannot be used when 2\,MB pages are exhausted, and once split into 2\,MB pages, they are difficult to recombine into 1\,GB pages. Moreover, cloud servers host diverse VM configurations not all aligned to 1\,GB boundaries. VM allocation depends on inventory scheduling, requiring flexible memory provisioning. Consequently, providers primarily reserve 2\,MB Hugetlb pages to maximize allocation flexibility, making 1\,GB pages difficult to allocate on demand.

\subsubsection{Inventory Management Challenges}

\captionsetup[figure]{skip=3pt}

\begin{figure}[htbp]
    \centering
    \begin{subfigure}[b]{0.48\linewidth}
        \centering
        \includegraphics[width=\linewidth,height=2.5cm]{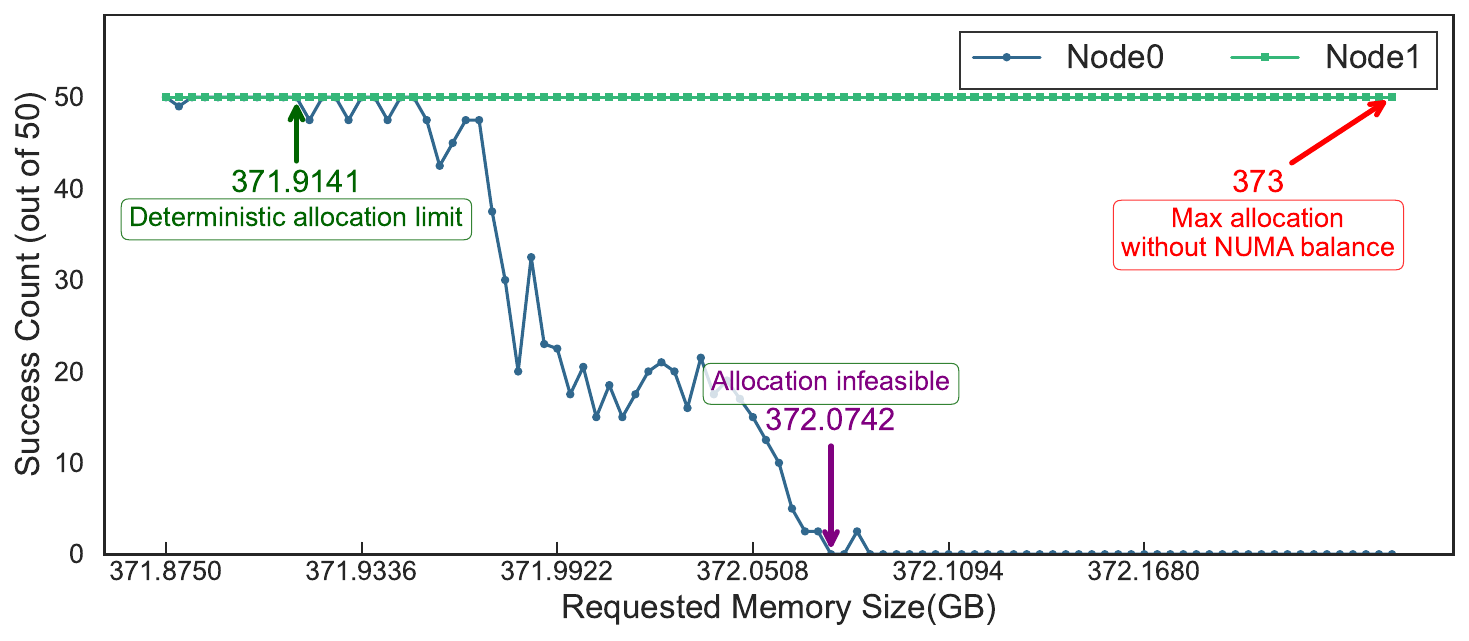}
        \caption{Allocation Success Rate.}
        \label{fig:Hugetlb_alloc}
    \end{subfigure}
    \hfill
    \begin{subfigure}[b]{0.48\linewidth}
        \centering
        \includegraphics[width=\linewidth,height=2.5cm]{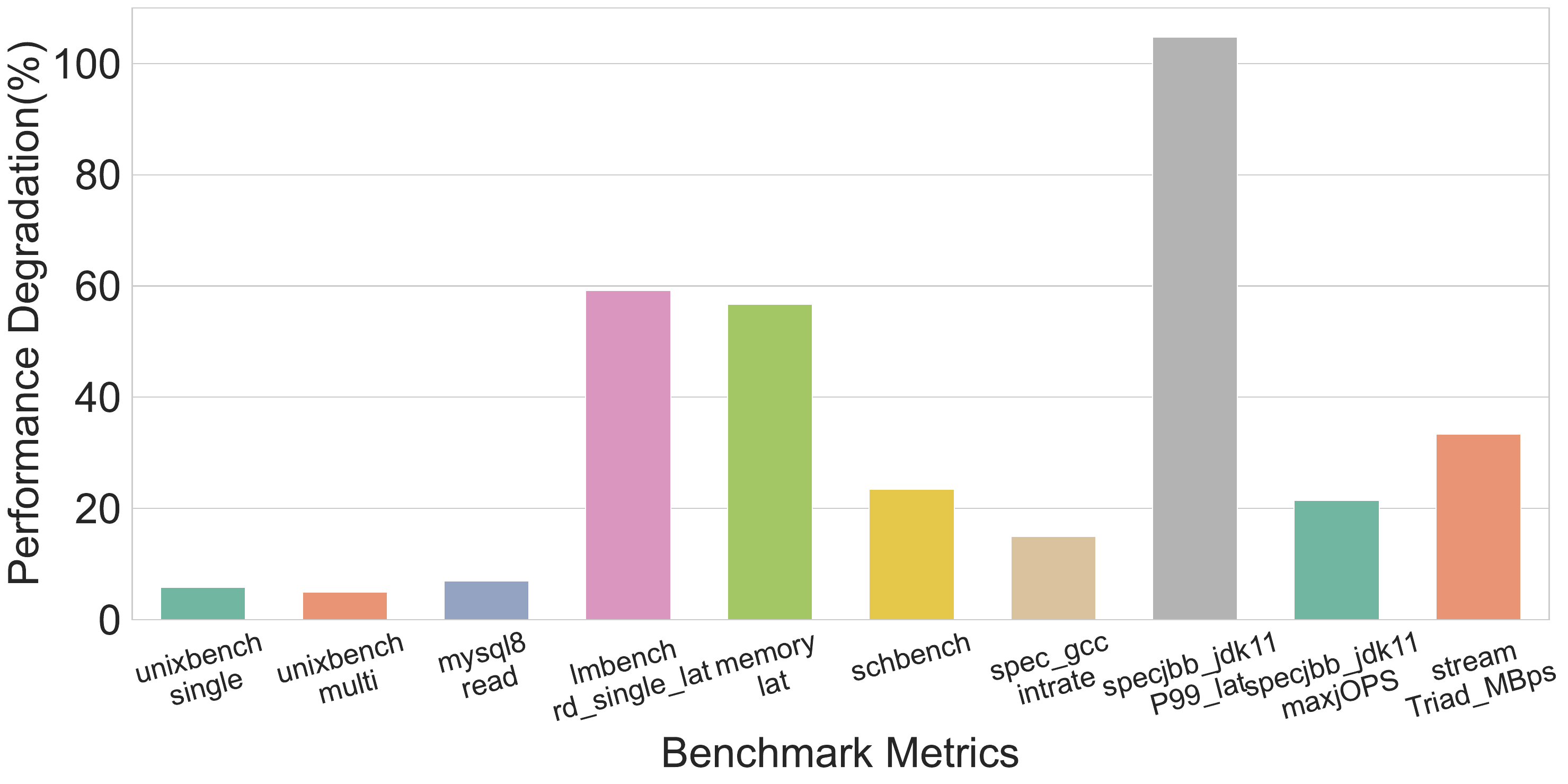}
        \caption{NUMA Imbalance Overhead.}
        \label{fig:numa}
    \end{subfigure}
    \caption{Hugetlb Inventory Challenges.}
    \label{fig:inventory}
\end{figure}

Due to memory fragmentation, Hugetlb cannot deterministically allocate maximum sellable memory, hindering production deployment. Even with boot-time pre-reservation, huge-page allocation is not guaranteed, requiring providers to cap sellable memory below theoretical limits. As shown in Figure~\ref{fig:Hugetlb_alloc}, tests on a 384\,GB 2-node server show that to reliably allocate 2\,MB Hugetlb pages under NUMA balance constraints, the total sellable memory must not exceed 371.91\,GB. Even below this threshold, occasional allocation failures occur; above it, failure rates rise sharply. Pre-reservation failures often require multiple reboots, increasing deployment time and unpredictability. In practice, maintaining NUMA balance is difficult beyond 372.07\,GB, and allocations beyond 373\,GB rarely succeed, even without balance constraints. Thus, under extreme allocation, Hugetlb cannot reliably guarantee deterministic allocation of the specified number of 2\,MB pages, and using 1\,GB pages of equivalent total size further reduces success rates. For either page size, allocation-failure probability increases with uptime, making runtime reconfiguration infeasible and limiting online inventory flexibility.

In modern cloud environments with multi-core processors and NUMA architectures, memory access locality greatly impacts VM performance. Cross-NUMA memory access can cause up to \texttt{100\%} performance degradation (Figure \ref{fig:numa}). Performance sensitive users often require VM-visible NUMA topology, necessitating NUMA-balanced physical memory allocation. To achieve this, VM memory must be evenly distributed across NUMA nodes for topology-aware scheduling. However, when VM memory shares space with host OS memory, kernel fragmentation disrupts balanced allocation, even with boot-time Hugetlb reservation. Under maximum allocation (Figure \ref{fig:Hugetlb_alloc}), node0 typically fragments earlier than node1, causing imbalanced sellable capacity. This results in unbalanced inventory: despite available memory, providers may be unable to sell due to the risk of NUMA imbalance and performance loss. Thus, inventory friendliness is critical to maximizing cloud memory utilization.

\subsubsection{Extreme Elasticity Challenges}
\label{fast_boot}
Cloud computing demands extreme elasticity, with rapid VM instantiation and termination. However, many VMs now employ passthrough devices, which require memory pinning at startup. While traditional Hugetlb supports prereserving huge-page pools, applications still rely on page faults to establish page-table mappings—a process that becomes prohibitively slow for large memory allocations. For VMs with pass-through devices, all assigned memory must be pinned at boot to prevent runtime migration, requiring complete traversal of the entire memory space before starting \cite{boot20}. As shown in Table \ref{tab:hugeltb_pagefault}, VMs of different sizes incur thousands of page faults during boot, with a 373GB VM experiencing over 35K page faults in the boot phase alone.

Additionally, the system must traverse page tables to collect memory layout and report contiguous regions to the VFIO driver for IOMMU mapping-further delaying boot. In practice, startup a 373\,GB VM with Hugetlb takes almost 100\,s, of which approximately 79\,s is spent setting the page-table based on faults and 13\,s bind to device memory. This memory preparation overhead scales with memory size: larger VMs incur more page faults, longer mapping times, and higher pinning latency. Eliminating this boot-time memory initialization would significantly accelerate the VM startup.

\begin{table}[t]
\caption{Hugetlb Page Faults Across Memory Sizes.}
\label{tab:hugeltb_pagefault}
\scriptsize
\setlength{\tabcolsep}{4pt}
\begin{tabular}{lrrrrrrr}
\toprule
\textbf{Memory (GB)} & 4 & 16 & 32 & 64 & 128 & 256 & 373 \\
\midrule
\textbf{Page Faults (K)}   & 1    & 4    & 9     & 12    & 17    & 21    & 35 \\
\textbf{Startup Time (s)}     & 10.24 & 11.66 & 14.54 & 19.56 & 31.52 & 48.61 & 100.12 \\

\bottomrule
\end{tabular}
\end{table}

\subsection{Opportunities for Cloud Memory Management}

\subsubsection{Limitations of Kernel-Reserved Memory}
Linux memory management includes a special \texttt{PFNMAP} type that does not require \texttt{struct page} metadata (green area in Figure \ref{fig:mm}). Reserving most physical memory upfront for VMs using \texttt{PFNMAP} eliminates \texttt{struct page} overhead, significantly reducing memory waste. However, as shown in Table \ref{tab:flow_pfnmap}, kernel support for \texttt{PFNMAP} is incomplete: it assumes contiguous memory by default, lacks native huge page page-table support, and suffers from slow page-table setup. These issues can be addressed with moderate, low-cost modifications. 

\begin{table}[htbp]
\caption{Current Flow of \texttt{PFNMAP} Memory.}
\label{tab:flow_pfnmap}
\scriptsize
\setlength{\tabcolsep}{3pt}
\begin{tabular}{lcccccc}
\toprule
& \textbf{Granularity} & \textbf{Layout} & \textbf{PT Mapping} & \textbf{PT Setup} & \textbf{Elastic} & \textbf{Faults} \\
\midrule
Current & Coarse & Contiguous & 4K & Slow & No & No \\
Desired & Fine & Flexible & 4K/2M/1G & Fast & Yes & Yes \\
\bottomrule
\end{tabular}
\end{table}

Moreover, in cloud environments, reserved memory must be shared among multiple VMs rather than used as a single block, requiring a additional memory manager. Fortunately, VM memory usage scenarios in cloud computing are limited, and relatively lightweight memory management suffices to meet these needs. Additionally, a key limitation of this reservation approach is that it fully isolates VM and OS memory pools, preventing resource sharing. When the OS experiences memory pressure, it cannot reclaim reserved memory—an essential capability for system stability. Without such elasticity, providers must conservatively reserve more memory for the OS, reducing sellable capacity and hurting utilization. Therefore, to gain the benefits of reserved memory while meeting cloud computing requirements, reserved memory must also support elastic provisioning.

\subsubsection{Limitations of Recent Huge-page Solutions}
Although Hugetlb is widely used in traditional cloud to provide memory for VMs, recent proposals \cite{folio, HVO, dmemfs} have introduced improvements mainly targeting metadata overhead and huge-page granularity. As shown in Table \ref{tab:large_page_comparison}, Folio provides more flexible huge-page sizes and optimizes locking and metadata efficiency for huge memory operations. However, in cloud scenarios, the 2MB granularity already offers sufficient inventory flexibility and strong performance, conclusion long validated in production. Huge pages smaller than 2MB provide no inventory benefits and require 4KB page-table mappings, which degrade performance. Thus, while Folio improves huge-page granularity flexibility, it brings no additional benefit in cloud environments.

HVO and Dmemfs significantly reduce metadata overhead, improving sellable memory. However, HVO does not fully eliminate \texttt{struct page} overhead—retaining \texttt{12.5\%} metadata cost for 2MB pages—and lacks broad production deployment. Dmemfs completely removes \texttt{struct page} overhead for VM memory, but neither supports elastic memory sharing between VM and host OS, requiring conservative OS memory reservation. Moreover, although both support multiple huge-page granularities, they use isolated pools that cannot be reconfigured at runtime, and still rely on page faults for allocation, failing to address slow VM startup.

\begin{table}[htbp]
\caption{Comparison of Modern Huge Page Techniques.}
\label{tab:large_page_comparison}
\scriptsize
\setlength{\tabcolsep}{3.5pt}
\begin{tabular}{lccccc}
\toprule
\textbf{Feature} & \textbf{Hugetlb} & \textbf{Folio} & \textbf{HVO} & \textbf{Dmemfs} & \textbf{Desired}\\
\midrule
Hot Upgrade          & No & No & No & No & Yes \\
Metadata Overhead    & High & High & Med. & Low & Low\\
Elastic Resource Pool      & No & No & No & No & Yes \\
Page Granularity  & 2M, 1G & $2^{n} \times \mathrm{4K}$ & 2M, 1G & 2M, 1G & 2M, 1G \\
Shared Page Pools      & No & No & No & No & Yes \\
Deterministic Alloc.       & No & No & No & Yes & Yes \\
Fast Boot         & No & No & No & No & Yes \\
Kernel Modifications & Built-in & Extensive & Extensive & Extensive & Minimal  \\
\bottomrule
\end{tabular}
\end{table}

Critically, these approaches require extensive OS modifications, increasing integration complexity and maintenance burden. They remain dependent on general-purpose memory management, inheriting its complexity. Most importantly, none support in-place upgrades: bug fixes or feature additions require VM restarts or host rollouts, limiting agility. When hot patches fail, repairs require rolling maintenance, forcing providers to minimize features for stability. This hinders the long-term evolution of core cloud components. Consequently, Hugetlb remains the most widely adopted and production-proven huge-page solution in the industry.

\subsection{Desired Cloud Memory Management System}
Therefore, to avoid the various issues of traditional memory management, we define the following objectives for a cloud-native memory management:

\textbf{O1 - Lightweight, cloud-specialized design}: Avoid the burden of general-purpose memory management, retaining only the functions needed in cloud computing, and achieves specialization without extensive host OS modifications, shedding the legacy of generic memory management.

\textbf{O2 - Maximized sellable memory}: Eliminate the overhead of traditional memory management metadata and ensure deterministic, NUMA-balanced memory distribution for sale. While maintaining stability, minimize host OS memory to maximize customer-available memory.

\textbf{O3 - Huge-page-oriented management}: Inherit Hugetlb performance benefits in cloud computing, while designing a flexible huge-page sharing mechanism for on-demand allocation at varying granularities and scopes, reducing fragmentation and aiding cloud server inventory management.

\textbf{O4 - Allocation-to-availability efficiency}: Fast page table mapping and bidirectional address translation reduce memory provisioning latency, eliminating memory preparation overhead on VM start/stop and meeting cloud users’ demands for extreme elasticity.

\textbf{O5 - Hot-upgrade capability}: To support rapid iteration and long-term evolution of cloud computing products, the new dedicated memory manager must support online hot-upgrade, enabling timely bug fixes and feature extensions during large-scale production deployment.

\section{Overview}

\begin{figure}[htbp] 
   \centering
   \includegraphics[width=0.9\linewidth]{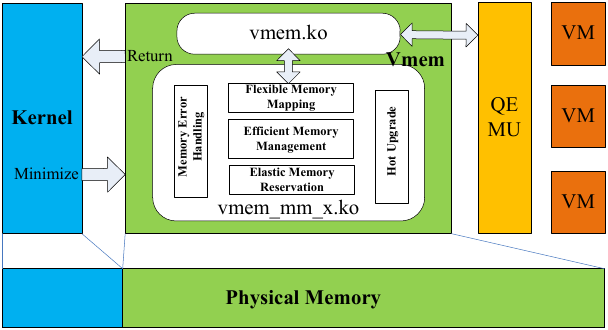}
   \caption{Vmem Architecture.} 
   \label{fig:arch} 
\end{figure}

As shown in Figure \ref{fig:arch}, Vmem is a cloud-based memory management system that integrates elastic reserved memory, memory slicing, mapping management, fault handling, and hot-upgrade capabilities.

First, Vmem enhances traditional OS reserved memory management with elastic reserved memory, decoupling VM memory from the host OS and eliminating VMs’ reliance on the \texttt{struct page} structure. It employs an inventory-friendly allocation method that compresses host OS memory usage, improves cloud server utilization, and allows reserved memory to be dynamically returned to the host OS for stability, achieving \textbf{O2}.  
Second, to avoid the complexity of traditional OS memory management, Vmem’s memory slice management is lightweight, using a novel allocation algorithm that allows huge pages of different granularities to share a unified reserved memory pool while minimizing fragmentation, fulfilling \textbf{O3}.  
Third, Vmem’s memory mapping management supports flexible, multi-granularity memory allocation and mapping, along with fast mapping mechanisms, to meet the performance and extreme elasticity needs of cloud workloads, accomplishing \textbf{O4}.  
Fourth, Vmem includes fault handling to address hardware memory errors such as Machine Check Exception (MCE), preventing use of faulty physical memory (this aspect is not the focus of this paper).  

Finally, Vmem uses a modular design to manage reserved memory, with all functionalities implemented as kernel modules independent of kernel memory management, avoiding extensive kernel modifications to meet cloud requirements and achieving \textbf{O1}. As a foundational cloud component, Vmem supports hot upgrades to ensure stability and maintainability. It separates memory management into two modules: the interface module \texttt{vmem.ko} and the core logic module \texttt{vmem\_mm.ko}. The lightweight \texttt{vmem.ko} exposes a unified interface via \texttt{/dev/vmem} and requires no hot upgrades, while \texttt{vmem\_mm.ko} implements the cloud-oriented memory management. Updates are applied by replacing \texttt{vmem\_mm.ko}, enabling seamless hot upgrades and fulfilling \textbf{O5}.

Through the \texttt{/dev/vmem} interface provided by \texttt{vmem.ko}, Vmem serves user-mode processes, while the core logic in \texttt{vmem\_mm.ko} can be iteratively updated transparently to users. Initially implemented on Linux-4.9 for KVM/QEMU to manage reserved VM memory, Vmem’s lightweight design and hot-upgrade capabilities ensure high portability. It has since been extended to Linux 4.19/5.10, x86, AMD, and ARM architectures, and adopted beyond VM scenarios (e.g., DPUs). Currently, Vmem has been a foundational component of our public cloud for over seven years, stably serving hundreds of millions of VMs.

\section{Memory Management}
\subsection{Memory Reservation}
Vmem’s memory reservation removes metadata overhead, ensures deterministic and evenly distributed sellable memory for upper-layer inventory management, and minimizes host OS usage without compromising stability, maximizing memory available to VMs.

\subsubsection{Balanced Reservation}
In cloud services, most memory is sold to VMs. Vmem accomplishes this through reserved memory: at system startup, all physical memory except that required by the host OS is assigned to Vmem. This removes VM memory metadata overhead, increases sellable memory, and fully isolates host OS and VM memory, ensuring reserved memory is deterministically sellable—unlike Hugetlb, where fragmentation can cause huge-page instability. However, deploying reserved memory in cloud scenarios requires more than simply increasing sellable memory; it must be provisioned according to inventory management needs to maximize capacity. NUMA distribution must be considered to avoid performance degradation or unsellable VMs from uneven inventory. To support inventory management in complex NUMA architectures and with physically non-contiguous memory, Vmem balances reserved memory across nodes so that each reserves an equal amount, preventing resource waste from inter-node imbalance.

\begin{figure}[htbp] 
   \centering
   \includegraphics[width=\linewidth]{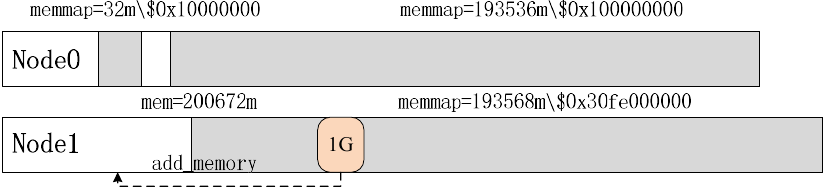}
   \caption{Balanced Multi-node Memory Reservation.} 
   \label{fig:multi-node} 
\end{figure}

As shown in Figure \ref{fig:multi-node} , for a host with two nodes and 384\,GB of memory, when Vmem reserves 378\,GB, balanced multi-node reservation is achieved by setting the \texttt{mem} and \texttt{memap} parameters. Each node reserves 193{,}586\,MB, of which 189\,GB is sellable memory and 32\,MB is for memory fault handling, facilitating cloud server inventory management. This balanced reservation method deterministically reserves NUMA-balanced sellable memory, avoiding the uncertainty and imbalance of Hugetlb maximum reservations and simplifying upper-layer inventory sales management.

\subsubsection{Elastic Reservation}
While reserved memory cleanly isolates host OS and VM memory for independent VM management, it alone cannot maximize cloud server memory sales. In traditional reservation, reserved memory cannot be shared with the host OS after boot. Given production environment complexity, ensuring sufficient host OS memory is difficult. Typically, a fixed empirical value—e.g., 6\,GB in the figure—is reserved, which suffices in most cases, but without a fallback, operators avoid reducing it to this minimum, instead reserving more for special cases. This extra memory often remains idle, reducing sellable capacity.

Vmem provides an elastic reserved-memory adjustment mechanism that dynamically returns unused Vmem-managed memory to the host OS when needed, using memory hotplug technology (as shown in Figure \ref{fig:multi-node}). This allows host OS memory to be tightly constrained in production; if insufficient, additional memory can be promptly returned, maximizing VM-allocatable memory while avoiding OOM errors or system crashes. Since Vmem selects the physical addresses of returned memory, the NUMA layout remains compliant with inventory requirements, enabling rapid online reconfiguration of VM memory without reboot and supporting elastic cloud inventory scheduling.

\subsection{Memory Slicing}
Since VM memory is used by the guest OS, the host’s primary memory-management task is allocating and reclaiming reserved memory. This entails slicing the large reserved space into smaller units for allocation at chosen granularities. As huge pages of different sizes provide varying virtualization benefits, the slicing mechanism must support sharing huge-page resources of multiple granularities.

\subsubsection{Slice Management}
To allocate large reserved memory blocks to VMs, Vmem slices all reserved memory at a fixed granularity. To meet lightweight metadata (\textbf{O1}), per-slice metadata (e.g., usage state) must be minimal. Vmem uses lightweight node abstractions (structures) and slice abstractions (arrays) to track the usage state of all nodes and their slices. Slicing granularity critically affects metadata overhead and complexity: overly fine granularity increases overhead and may degrade VM performance, while overly coarse granularity reduces allocation flexibility. Based on Section~\ref{huge_perf} and empirical experience with VM inventory management and huge-page performance in production, a 2\,MB slice granularity balances performance and flexibility.

\begin{figure}[htbp] 
   \centering
   \includegraphics[width=\linewidth, height=3.5cm]{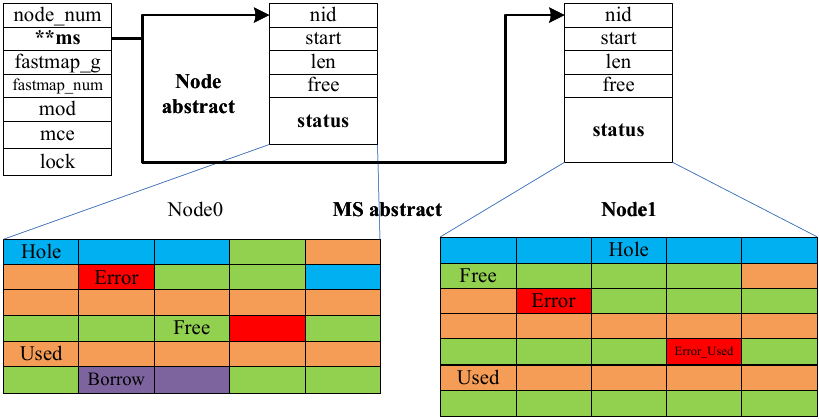}
   \caption{Slice Structures of Vmem.} 
   \label{fig:ms} 
\end{figure}

In cloud computing, VM memory usage scenarios are limited but varied, requiring only a finite set of states. Bitmap-based tracking records only allocation and free states; extending it for memory faults or returning memory to the OS needs extra bitmaps, increasing complexity. To balance extensibility, complexity, and overhead, Vmem stores each slice’s state in a 1-byte \texttt{char}. As shown in Figure \ref{fig:ms}, each node has a memory section (\texttt{ms}) structure, and since reserved memory is physically contiguous, an array suffices to track slice states within a node. States include \texttt{free}, \texttt{used}, \texttt{hole}, \texttt{error}, \texttt{mce}, \texttt{mce\_used}, and \texttt{borrow}. In a 384\,GB cloud server, this design needs only 192\,KB for state management—negligible in practice—while reducing metadata overhead, supporting diverse scenarios, and preserving extensibility.

\subsubsection{Slice Allocation}
Considering the performance benefits of huge pages for VMs, Vmem preferentially allocates memory slices using huge pages. Users can specify the size, allocation mode, and huge-page granularity \texttt{psize}, with allocation start addresses aligned to \texttt{psize}. Supported granularities are \texttt{1GB}, \texttt{2MB}, and \texttt{mix}, where \texttt{mix} prefers 1\,GB and falls back to 2\,MB. While 2\,MB slices provide substantial performance in most cases, some scenarios benefit from 1\,GB pages, so Vmem shuold supports both granularities from the outset for on-demand sharing.

Although Vmem avoids 2\,MB huge page fragmentation, it still experiences 1\,GB fragmentation, with aligned segments harder to find over time. Conventional methods—separate pools \cite{litke2007turning,lu2006using}, extra tracking structures \cite{dmemfs}, or buddy-like mechanisms—are complex \cite{huang2016evolutionary,park2018gcma}, add overhead, and compromise Vmem’s lightweight design. To leverage both huge page sizes while avoiding fragmentation, Vmem adopts a mixed huge page bidirectional allocation strategy: allocate 1\,GB-aligned memory forward and 2\,MB-aligned memory backward, meeting in the middle to consume all free memory. To maintain 1\,GB contiguity, Vmem applies the following allocation policy:
\textbf{1)} Allocate 1\,GB-aligned blocks first; when 1\,GB alignment is not possible, allocate all remaining memory with 2\,MB alignment.  
\textbf{2)} For 2\,MB allocations, preferentially use fragmented blocks (where a 1\,GB-aligned region is partially used and no longer meets 1\,GB allocation requirements).  
\textbf{3)} Only when no fragmented free blocks remain may 2\,MB allocations use a fully free 1\,GB-aligned block.

\begin{figure}[htbp] 
   \centering
   \includegraphics[width=\linewidth, height=3.5cm]{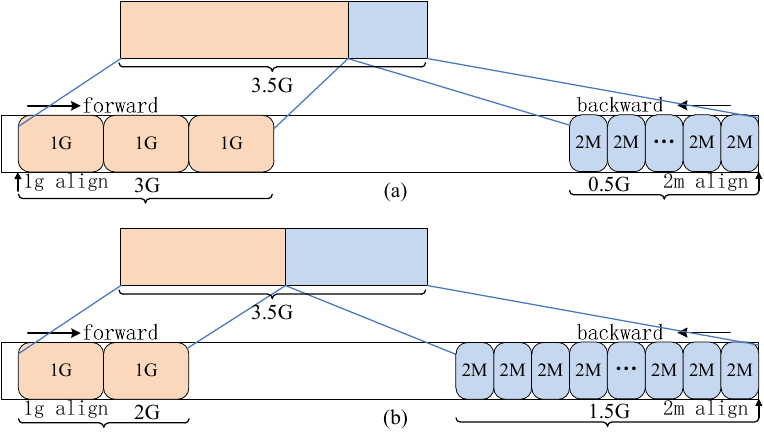}
   \caption{Bidirectional Memory Allocation Strategy.} 
   \label{fig:toward} 
\end{figure}

As shown in Figure \ref{fig:toward} , a request of \texttt{size} is split into \texttt{size\_1G} and \texttt{size\_2M}, representing portions allocated with 1\,GB and 2\,MB alignment, respectively. For example, for a 3.5\,GB request, \texttt{size\_1G} = 3\,GB and \texttt{size\_2M} = 0.5\,GB (Figure \ref{fig:toward}a). If forward allocation finds only 2\,GB of contiguous 1\,GB-aligned memory, then \texttt{size\_1G} = 2\,GB and \texttt{size\_2M} = 1.5\,GB (Figure \ref{fig:toward}b). In all cases, \texttt{size} = \texttt{size\_1G} + \texttt{size\_2M}, with the division determined by the current memory state.
With memory-slice management, Vmem flexibly handles different granularities without Hugetlb’s non-sharing limitation, avoiding complex overhead while greatly improving cloud server memory allocation flexibility for diverse scenarios. For cases not requiring preallocated memory (e.g., VMs without passthrough devices), Vmem also supports on-demand allocation via page faults.

\subsection{Memory Mapping}
To meet extreme VM performance demands, Vmem’s page-table mapping fully exploits huge-page advantages. For cloud elasticity, especially fast starts, Vmem-allocated memory must be immediately usable, requiring a balance of functional flexibility, mapping efficiency, and translation efficiency.

\subsubsection{Mixed Page Table Mapping}

\begin{figure}[htbp] 
   \centering
   \includegraphics[width=\linewidth]{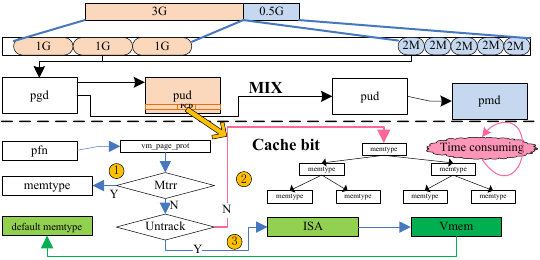}
   \caption{Page Table Map of Vmem.} 
   \label{fig:map} 
\end{figure}

Vmem allocates physical memory and maps it via page tables for VM use. However, current PFNMAP support lacks huge-page mapping and suffers from slow mapping (Table \ref{tab:flow_pfnmap}). To make reserved memory truly effective in cloud scenarios, these issues must first be addressed. Firstly, the kernel’s PFNMAP mechanism must be enhanced to support huge-page reserved memory. In cloud environments, different workloads may prefer 2MB or 1GB pages for performance. To maximize the benefits of huge pages—such as reduced TLB pressure and lower metadata overhead—Vmem supports multiple mapping granularities: \texttt{2MB}, \texttt{1GB}, and \texttt{mix}. For \texttt{2MB} or \texttt{1GB} mappings, Vmem establishes huge-page entries at the PMD- or PUD- level, respectively. For non-1GB-aligned allocations, \texttt{mix} mode prioritizes 1GB pages and falls back to 2MB, maximizing huge-page utilization (Figure~\ref{fig:map}). 

In addition, to enable fast memory allocation and rapid VM startup, efficient page-table setup is critical. However, the native kernel builds page tables for \texttt{PFNMAP} reserved memory slowly. As shown in Figure \ref{fig:map}, page-table creation requires setting the page’s cache flag, usually determined from the process \texttt{vma} structure’s \texttt{vma\_prot} (\orangecircled{1}). If the range is unrecognized by the kernel, it checks whether it is untracked memory and returns the default cache type (currently only for ISA, \orangecircled{3}). Otherwise, it takes the slow path, creating a \texttt{memtype} structure to record the desired cache type and range, inserting it into a red-black tree for later lookup (\orangecircled{2}). \texttt{PFNMAP} reserved memory currently follows this slow path, so each page-table entry creation incurs a red-black-tree lookup, making mapping slow and increasingly costly with larger memory sizes. To remove this overhead for large VMs, Vmem adds reserved memory to the untracked list to avoid slow-path lookups and accelerate page-table construction, significantly reducing page-table build costs and speeding VM startup.

\subsubsection{FastMap Bidirectional Address Translation}
\label{fastmap}
\begin{figure}[htbp] 
   \centering
   \includegraphics[width=\linewidth]{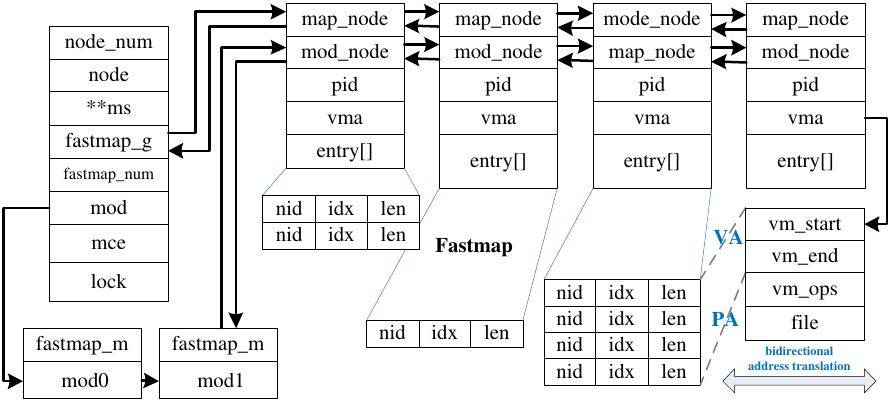}
   \caption{The FastMap for Vmem.} 
   \label{fig:fastmap} 
\end{figure}

During memory use, it is often necessary to query the mapping between physical and virtual memory, such as VM start/stop scenarios. The conventional method traverses page tables for address translation, which is costly. Since VMs typically use large amounts of memory and Vmem allocates it as contiguously as possible, physical memory is divided into only a few contiguous segments at runtime. Thus, a VM’s virtual-to-physical translation can often be achieved with a small number of linear mappings. As shown in Figure \ref{fig:fastmap} , Vmem constructs a \texttt{fastmap} structure recording allocation information, ending with an \texttt{entry} array storing all mapped physical memory regions of the VM. Each entry records the node, start PFN, size, and other details of a segment. \texttt{fastmap} also stores the VM’s \texttt{pid} and the \texttt{vma}. This lightweight structure enables fast bidirectional translation between a VM’s virtual memory and host's physical memory.

Since each VM issues only a limited number of \texttt{mmap} memory requests (typically once, or multiple for multi-nodes), the number of corresponding \texttt{fastmap} instances is also small. This design enables fast address translation and easy retrieval of contiguous memory regions, avoiding costly page-table traversal to enumerate all memory and greatly accelerating VM startup (see Section \ref{fast_boot}). It can also satisfy requirements such as zeroing memory for on-demand allocation and handling memory faults. During Vmem hot upgrades, the \texttt{vm\_ops} object of each \texttt{vma} that has opened the Vmem device, and the \texttt{fops} object of its associated file, must be updated. With the information recorded in \texttt{fastmap}, all operation structures for open Vmem devices can be located directly without traversing every \texttt{vma} of each VM process, enabling efficient interface pointer updates (see Section \ref{hotup}).

\section{Hot Upgrade}
\label{hotup}
\begin{figure}[t] 
   \centering
   \includegraphics[width=\linewidth]{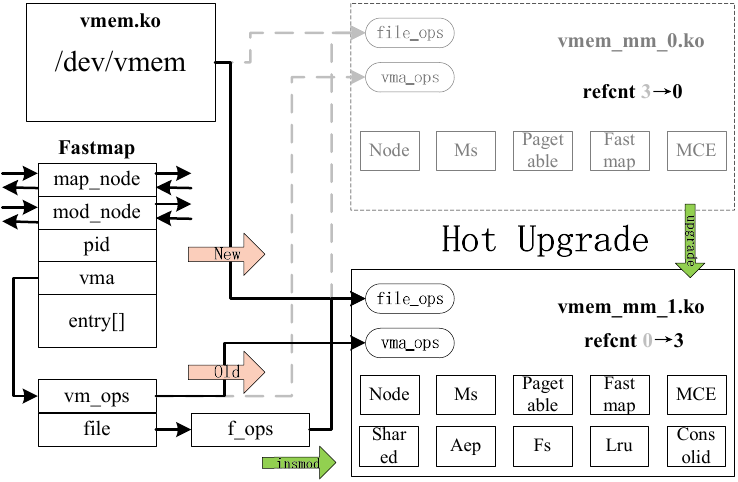}
   \caption{The Hot Upgrade for Vmem.} 
   \label{fig:upgrade} 
\end{figure}

Currently, no industry solution for hot-upgrading memory management functions. To enable iterative updates, Vmem is implemented as a kernel module decoupled from native OS memory management and split into two relatively independent modules. The interface module \texttt{vmem.ko} provides the memory management entry via the \texttt{/dev/vmem} character device, while \texttt{vmem\_mm\_[x].ko} (where \texttt{x} is the version) implements lightweight memory management logic. As shown in Figure \ref{fig:upgrade}, the interface module contains only basic interface functions and requires no hot upgrade, the core module, however, must be upgraded online. The Vmem hot-upgrade process proceeds as follows: 

\textbf{First}, the interface module exposes the \texttt{/dev/vmem} character device to VM management processes (e.g., QEMU), enabling allocation and release of VM memory. The device’s \texttt{cdev.ops} contains a set of \texttt{file\_operations} function pointers—such as \texttt{open}, \texttt{close}, \texttt{mmap}, and \texttt{ioctl}—that dispatch memory management operations to the core module. During an upgrade, only the new module is loaded, and the function pointers in \texttt{cdev.ops} are redirected to the updated implementations. However, the Linux kernel’s module reference counting (\texttt{refcnt}) mechanism prevents unloading of \texttt{vmem\_mm.ko} while any VM is active. If the old module contains a bug, it still poses a serious risk to running VMs, and keeping multiple memory management modules loaded is undesirable for online maintenance. 

\textbf{Second}, to enable safe removal, \texttt{refcnt} must be reduced to zero—requiring identification and decoupling of all runtime dependencies. Such dependencies mainly stem from operation functions in certain data structures still pointing to the old module. For example, memory allocated by Vmem is freed via the \texttt{close} path in the \texttt{vma\_operations} of the \texttt{vma}. If the new module’s memory release method is compatible, the function pointer for freeing already-allocated memory can be updated to the new module’s function. Similarly, the \texttt{f\_ops} field in the \texttt{file} associated with the \texttt{vma} also points to the old module’s \texttt{file\_operations}. To unload the old module, these operation fields in the \texttt{vma} must be updated. Vmem records all associated \texttt{vma} (Section \ref{fastmap}), and during pointer replacement, the old module’s \texttt{refcnt} is transferred to the new module. As the new module’s \texttt{refcnt} increases, the old module’s \texttt{refcnt} decreases; once it reaches zero, the old module can be safely removed and freed.

\textbf{Third}, this operation-set decoupling requires maintaining compatibility of Vmem’s metadata structures, which are directly inherited between old and new modules. When parsing shared data structures, the new module inherits the old module’s definitions. If the new module extends these structures and intends to reuse the old ones directly, extensions must use reserved fields to avoid parsing errors. Otherwise, dedicated parsing and conversion functions are needed to handle structural differences, but as online versions iterate, maintaining such incompatible conversions becomes costly and impractical. Ensuring data compatibility limits flexibility in extending metadata but avoids the complexity of maintaining conversions across versions.

\textbf{In addition}, during hot upgrades, the externally exported \texttt{/proc} interfaces must be updated. While an incremental approach similar to metadata compatibility could be used, the \texttt{/proc} interface often grows more complex and dependent on internal functions, making dependency decoupling cumbersome. As the outermost interface with short-lived operations and no dependent structures, it can be directly reconstructed by first unregistering the old module’s \texttt{/proc} entries and creating the new ones. Since the \texttt{/proc} filesystem uses an unload-and-load mechanism, entries can be freely changed without compatibility issues, offering greater flexibility.

\textbf{Finally}, like \texttt{file\_operations}, Vmem may export functions for use by the OS (e.g., the page-table mapping acceleration function) or other kernel modules (e.g., the VM startup acceleration function in \texttt{vfio}). These must be replaced with the corresponding functions from the new module. As with interface function replacement, if an exported function from the old module is executing, the update must wait for completion, ensured via RCU locks.  
With these steps, Vmem remains stable and efficient during hot upgrades. The basic memory management functions described here can then evolve online to support extended features. To reduce operational complexity, only two \texttt{vmem\_mm\_x} modules (e.g., \texttt{vmem\_mm\_0} and \texttt{vmem\_mm\_1}) are maintained, and different versions are hot-upgraded by switching between them.

\section{Evaluation and Experiment}
Vmem’s lightweight design improves memory utilization and cloud compatibility (Section~\ref{light}). Performance analysis shows its slicing and allocation fully exploit huge-page benefits (Section~\ref{perf}). Fast startup/shutdown enables instant availability and rapid mapping for extreme elasticity (Section~\ref{elast}), and hot-upgrade evaluation confirms architectural stability and efficiency (Section~\ref{hot}).

\subsection{Lightweight}
\label{light}
\subsubsection{Metadata Overhead Analysis}
In a production environment, the runtime metadata overhead of Vmem (Table~\ref{tab:meta}) comprises: 
(1) \textbf{Module Data}: \texttt{vmem} and \texttt{vmem\_mm} modules, fixed at 236\,KB unless updated.  
(2) \textbf{Basic MetaData}: core structures, including slice state (\texttt{vmem\_ms}), VM mapping (\texttt{vmem\_fastmap}), and fault handling (\texttt{vmem\_mce}). \texttt{vmem\_ms} scales with host size and slice granularity; \texttt{vmem\_fastmap} with contiguous-slice count. On a 2-node 384\,GB host with 2\,MB slices, \texttt{vmem\_ms} is 193\,KB; worst-case non-contiguous allocation yields 4608\,KB for \texttt{vmem\_fastmap}, while typical contiguous allocations need only 72\,B for a single-VM full-memory case or $\sim$7\,KB for fully loaded 2-core 4\,GB VMs.  
(3) \textbf{Other MetaData}: \texttt{/proc} filesystem, dump, and immutable data, fixed at 1760\,B.  
Overall, worst-case usage is $\sim$5039\,KB, and in a realistic 2-core 4\,GB deployment, $\sim$438\,KB. By contrast, HVO needs 0.75\,GB for 2\,MB huge pages, and dmemfs uses only tens of KB but lacks cloud-required capabilities. Given GB-scale host reservations, KB-level differences are negligible, while Vmem offers far more complete capabilities.

\begin{table}[t]
\caption{Detailed Metadata Overhead of Vmem.}
\label{tab:meta}
\small
\setlength{\tabcolsep}{5pt}
\begin{tabular}{lll}
\toprule
\textbf{Category}       & \textbf{Component}     & \textbf{Size (B)} \\
\midrule
\multirow{2}{*}{Module data} 
  & vmem       & 16,384 \\
  & vmem\_mm   & 225,280 \\
\midrule
\multirow{4}{*}{Basic metadata} 
  & vmem\_ms        & $112 \times \text{nodes} + \text{slices}$ \\
  & vmem\_fastmap   & $120 \times \text{nodes} + 24 \times \text{entry}$ \\
  & vmem\_mce       & $8 + 24 \times 8 \times \text{mce}$ \\
\midrule
\multirow{2}{*}{Other metadata} 
  & vmem\_proc      & 224 \\
  & vmem\_dump\_para & 16 \\
  & other immutable & 1,520 \\
\bottomrule
\end{tabular}
\end{table}

\subsubsection{Code Analysis}

As shown in Table \ref{tab:code}, Vmem’s codebase spans 15,747 lines, significantly lighter than Linux 4.9’s 128k+ lines for memory management. Core features (slicing, allocation, mapping) require only thousands of lines, with hot upgrades implemented in 353 lines due to modular design and fastmap. The code size demonstrates Vmem’s lightweight and efficient design, using minimal code to support large-scale cloud computing workloads in production.

\begin{table}[t]
\caption{Code Distribution of Vmem (total: 15,747 lines).}
\label{tab:code}
\scriptsize
\setlength{\tabcolsep}{12pt}
\begin{tabular}{llr}
\toprule
\textbf{Component} & \textbf{Functionality} & \textbf{Lines} \\
\midrule
vmem          & Interface device                    & 1,785 \\
vmem\_mm      & Common functionality                &   784 \\
vmem\_mm\_node & Elastic reserved memory             &   349 \\
vmem\_ms      & Memory slicing                      & 2,602 \\
vmem\_map     & Page table and fast mapping         & 3,670 \\
vmem\_mce     & Error handling                      &   786 \\
vmem\_hotup   & Hot upgrade support                 &   353 \\
vmem\_info    & Version, dump, procfs               & 2,899 \\
other codes   & Auxiliary functions                 & 2,519 \\
\midrule
\textbf{Sum}           & \textbf{Total}                               & \textbf{15,747} \\
\bottomrule
\end{tabular}
\end{table}

\subsubsection{Adaptation Analysis}
Enabling Vmem in cloud VMs requires adaptations across all virtualization layers including QEMU, libvirt, VFIO, and the kernel. Table \ref{tab:adapt} shows the code changes for each layer on x86 with the Linux 4.9 kernel. Most changes involve detecting and supporting reserved huge pages. Kernel modifications are comparatively larger, primarily to backport 1GB huge page support from newer kernels. Other layers require only minor adaptations to replace the original OS memory management with Vmem.

\begin{table}[t]
\caption{Vmem Adaptation Across Virtualization Layers.}
\label{tab:adapt}
\scriptsize
\setlength{\tabcolsep}{10pt}
\begin{tabular}{lcccccc}
\toprule
         & \textbf{QEMU} & \textbf{Libvirt}  & \textbf{VFIO} & \textbf{KVM} & \textbf{Kernel} \\
\midrule
LoC       & 145  & 59      & 372  & 52  & 563   \\
\bottomrule
\end{tabular}
\end{table}

\begin{figure*}[htbp]
\centering
\begin{subfigure}[b]{0.48\linewidth}
   \centering
   \includegraphics[width=0.9\linewidth]{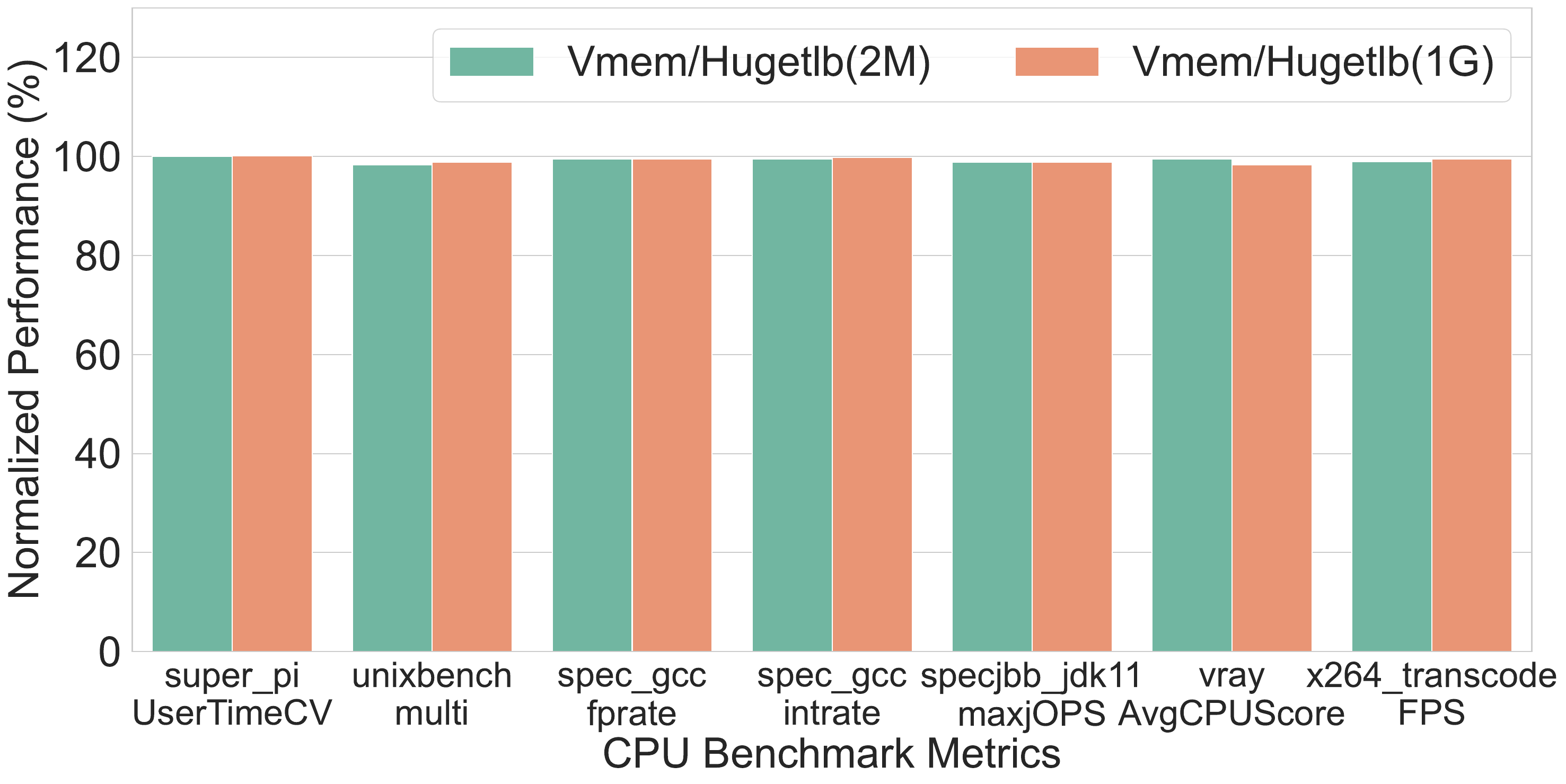}
   \caption{CPU}
   \label{fig:cpu}
\end{subfigure}
\hfill
\begin{subfigure}[b]{0.48\linewidth}
   \centering
   \includegraphics[width=0.9\linewidth]{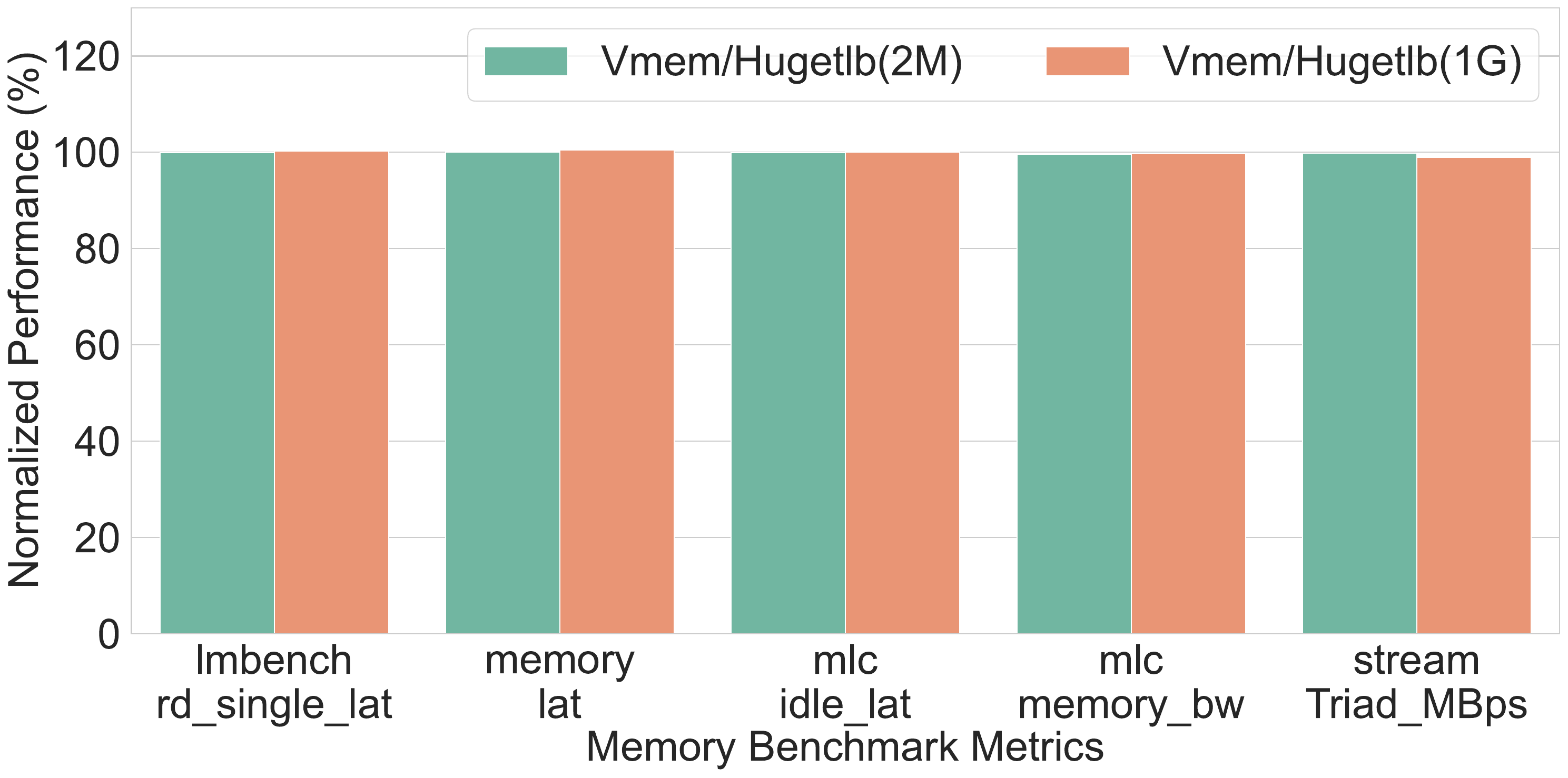}
   \caption{Memory}
   \label{fig:mem}
\end{subfigure}

\vspace{10pt} 

\begin{subfigure}[b]{0.48\linewidth}
   \centering
   \includegraphics[width=0.9\linewidth]{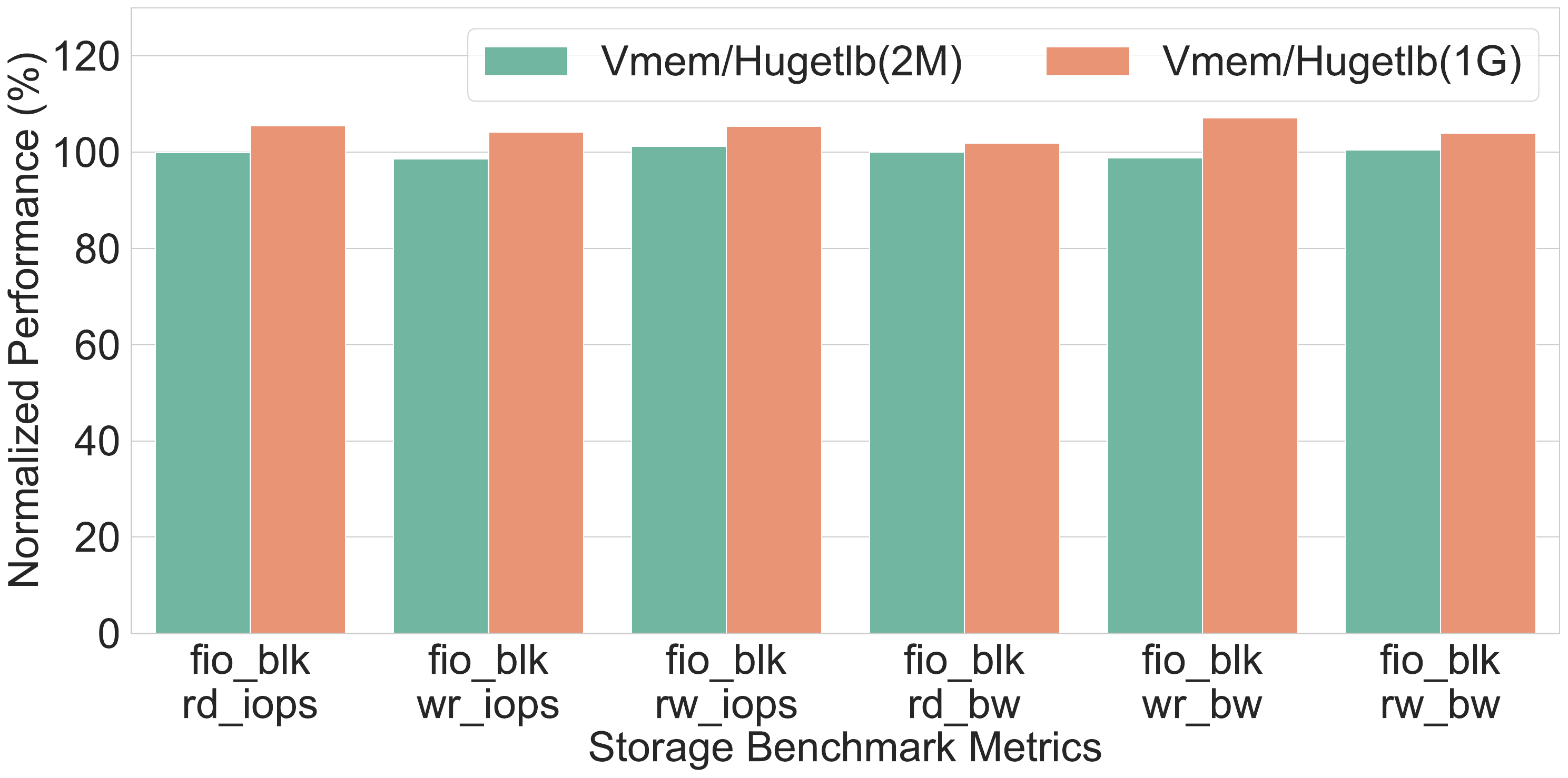}
   \caption{Storage}
   \label{fig:storage}
\end{subfigure}
\hfill
\begin{subfigure}[b]{0.48\linewidth}
   \centering
   \includegraphics[width=0.9\linewidth]{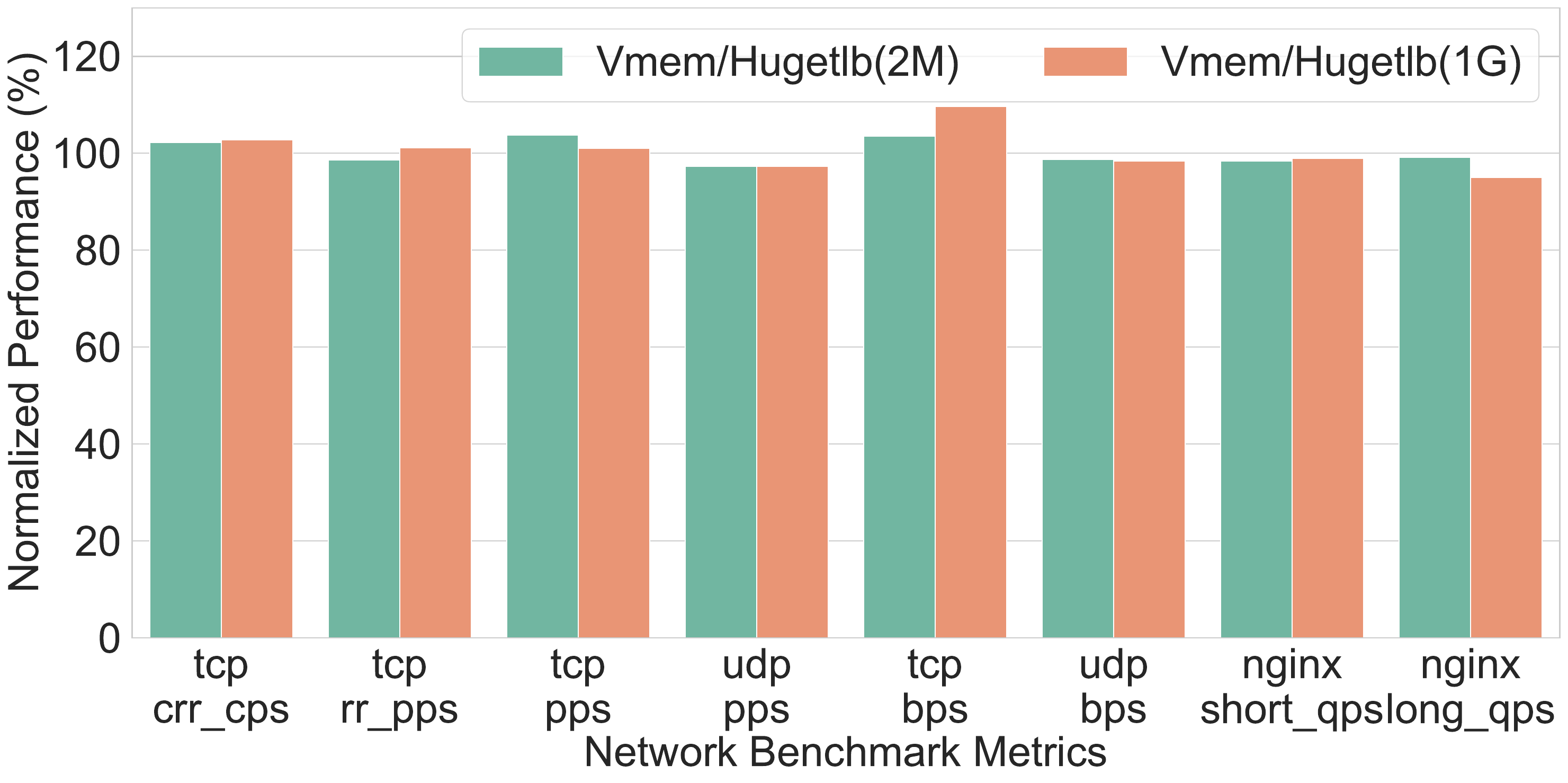}
   \caption{Network}
   \label{fig:network}
\end{subfigure}
\caption{Performance Under Diverse Workloads.}
\label{fig:overall1}
\end{figure*} 

\subsection{Performance}
\label{perf}

We evaluated Vmem on cloud servers (Intel® CPU @ 2.50\,GHz, 104 CPUs, 384\,GB RAM) against Hugetlb under identical kernels and virtualization components. As shown in Figure~\ref{fig:overall1}, workloads included computation, memory, disk I/O, and networking. Vmem matches Hugetlb in CPU and memory performance, as both use huge-page mappings at runtime and differences in allocation and release do not affect execution behavior. Disk bandwidth and IOPS across read/write shows no degradation, and network performance matches or exceeds Hugetlb. Vmem also exhibits no significant performance difference between 1\,GB and 2\,MB page mappings, due to ample TLB resources. Overall, Vmem delivers Hugetlb-level performance across all scenarios while improving memory utilization.

\subsection{Elasticity}
\label{elast}
\begin{figure}[t] 
   \centering
       \includegraphics[width=\linewidth, keepaspectratio]{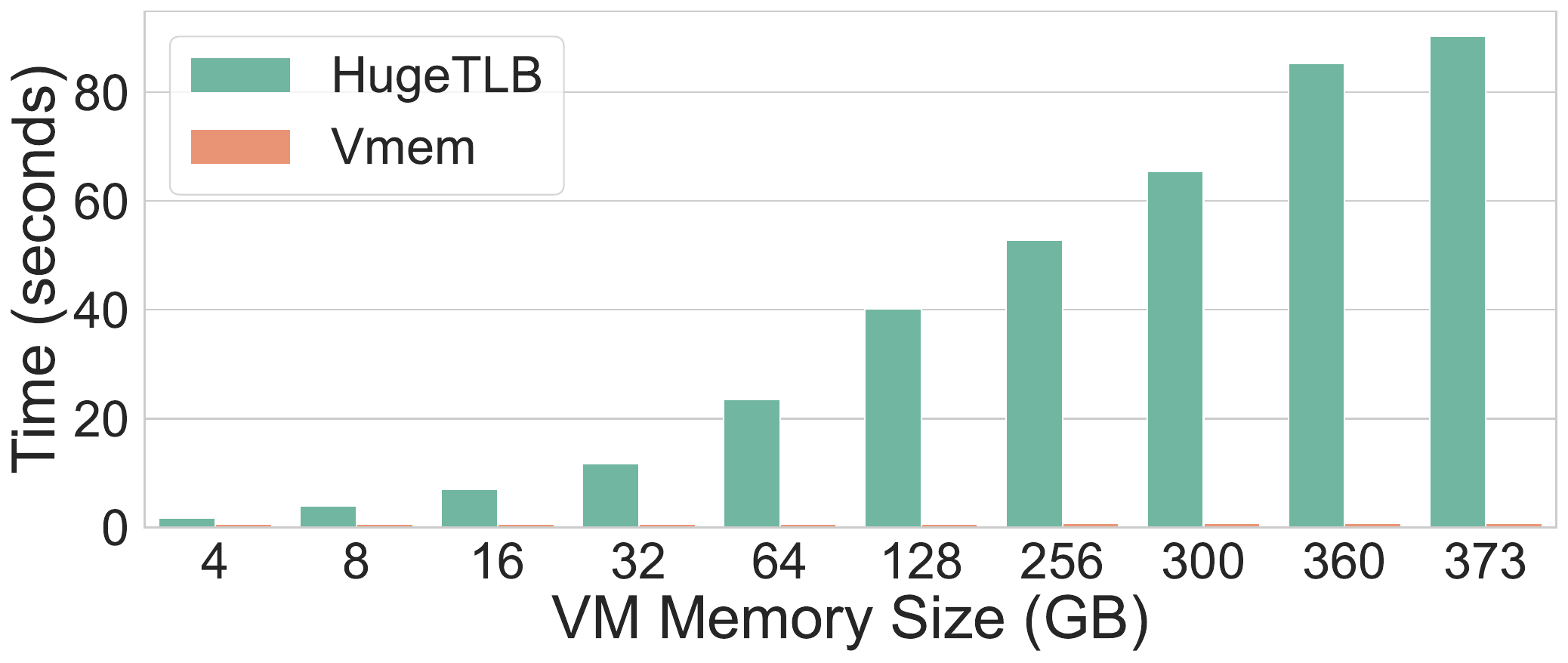}
   \caption{Virtual Machine Creation Time Comparison.} 
   \label{fig:start} 
\end{figure}

\begin{figure}[t] 
   \centering
    \includegraphics[width=\linewidth, keepaspectratio]{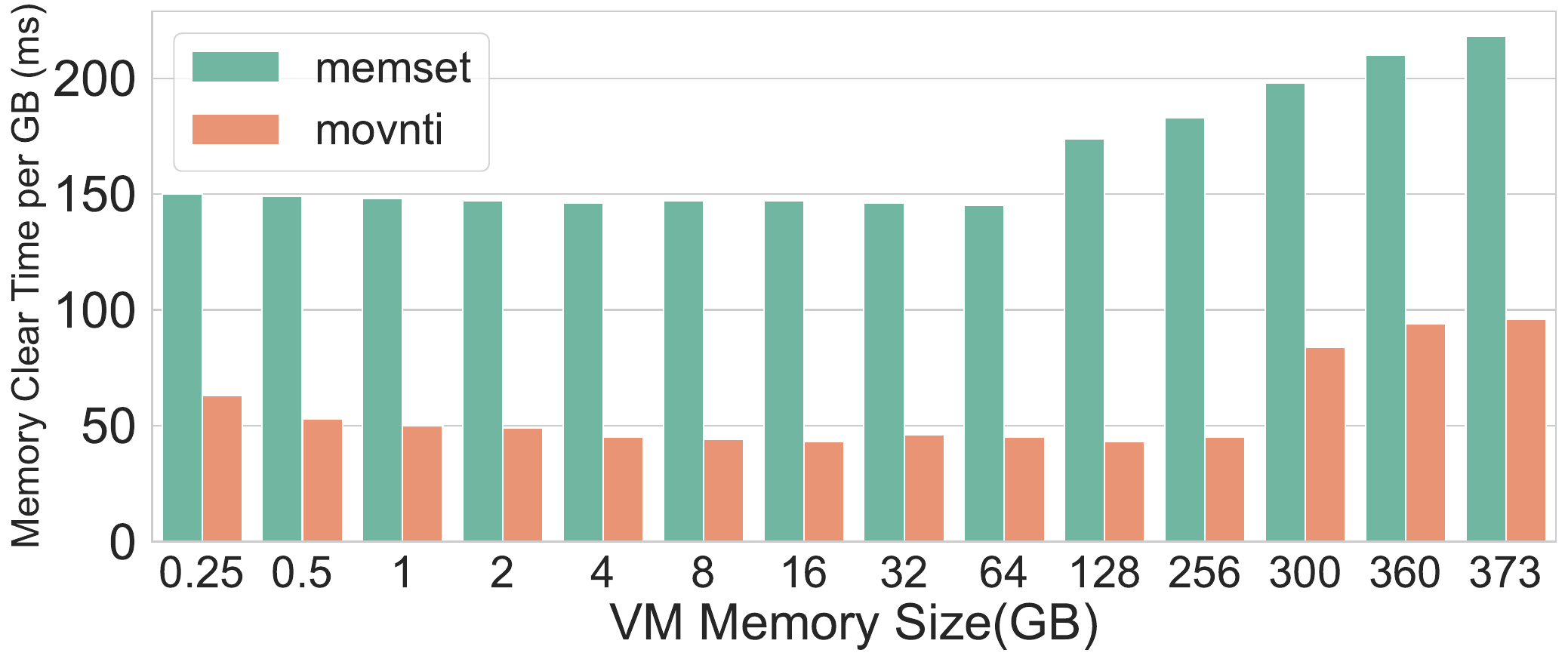}
   \caption{Memory Clear Time Comparison.} 
   \label{fig:clean} 
\end{figure}

To evaluate the performance benefits of Vmem's reserved memory mechanism for VFIO-passthrough VMs\cite{zhang2024hd} in cloud environments, we measured the creation times of VMs with varying memory sizes, comparing Vmem against traditional Hugetlb. As shown in Figure~\ref{fig:start}, Hugetlb creation time grows linearly with memory size, reaching 90\,s at 373\,GB. Vmem’s fast mapping and address translation enable much faster booting, achieving an average of 0.6\,s with creation time nearly constant across memory sizes. Furthermore, in multi-tenant cloud environments, memory zeroing is essential to prevent data leakage \cite{chatzoglou2024keep}. While allocation-time zeroing delays VM creation, Vmem’s fast startup aligns well with shutdown-time zeroing. Figure~\ref{fig:clean} compares Vmem’s \texttt{movnti}-based zeroing with traditional \texttt{memset} across VMs of varying memory sizes. \texttt{movnti} significantly outperforms \texttt{memset} by bypassing cache flushes, with both methods speeding up until 4\,GB (bandwidth saturation) and declining beyond 128\,GB due to NUMA latency. Thus, Vmem delivers strong Creation and shutdown performance, meeting cloud’s extreme elasticity requirements for rapid provisioning and deprovisioning.

\subsection{Hot Upgrade}
\label{hot}
\begin{figure}[t]
    \centering
    \begin{subfigure}[b]{\linewidth}
        \centering
        \includegraphics[width=\linewidth, height=5cm, keepaspectratio]{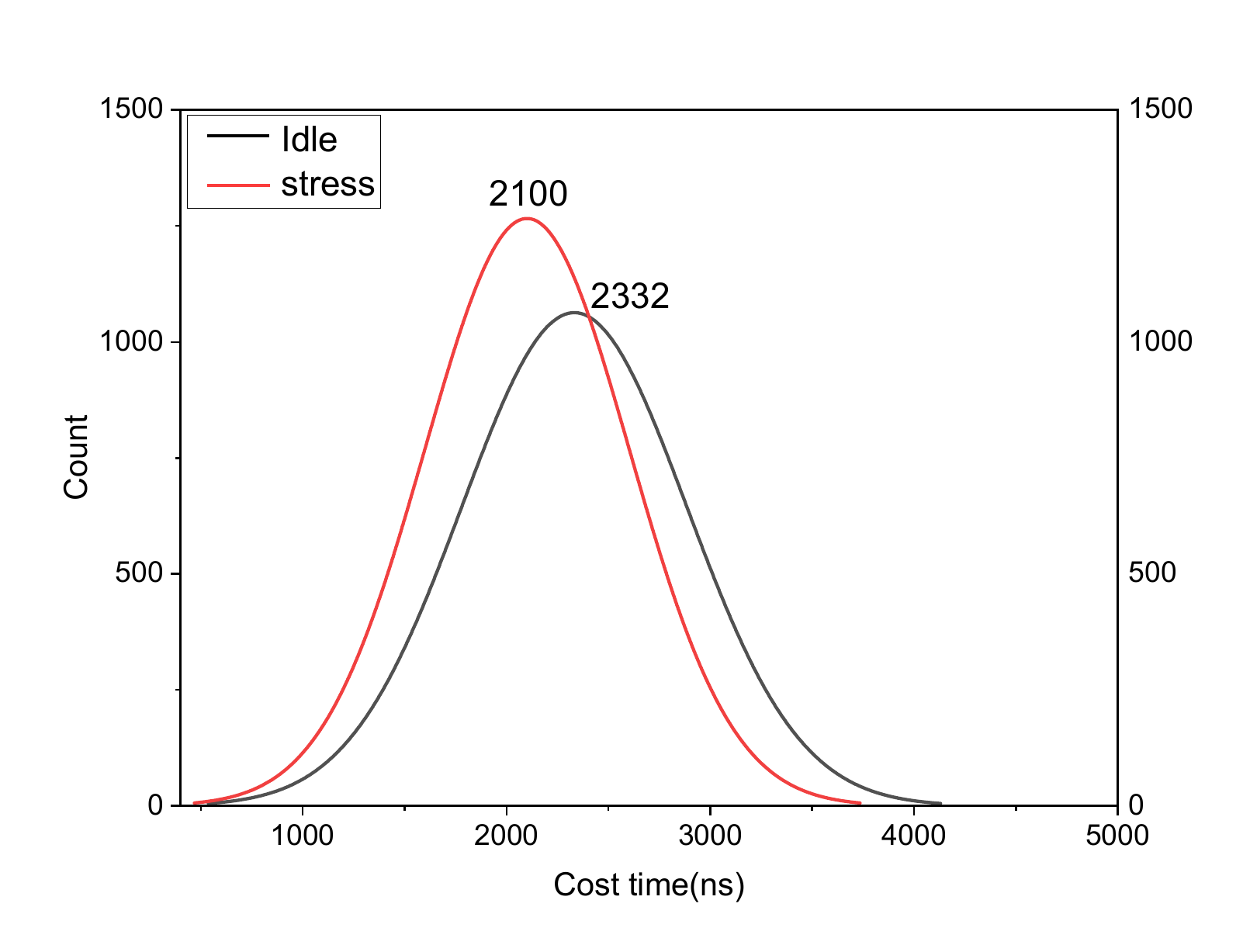}
        \caption{The Time Overhead.}
        \label{fig:upgrade_time}
    \end{subfigure}
    
    \begin{subfigure}[b]{\linewidth}
        \centering
        \includegraphics[width=\linewidth, height=5cm, keepaspectratio]{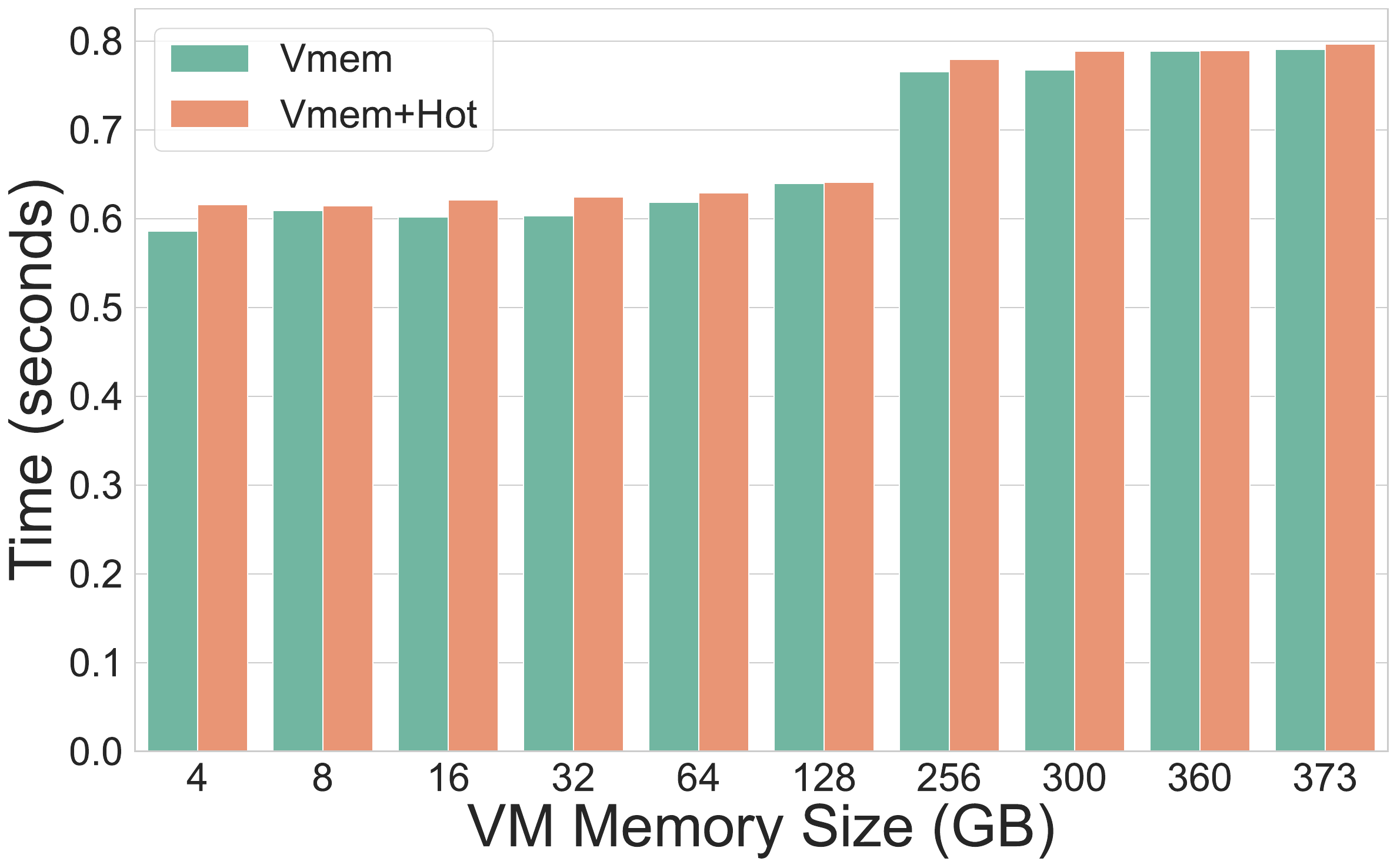}
        \caption{Creation Time Comparison.}
        \label{fig:upgrade_start}
    \end{subfigure}
    
    \caption{The Overhead of Hot Upgrade.}
    \label{fig:upgrade_data}
\end{figure}

Figure~\ref{fig:upgrade_time} shows the time overhead of Vmem’s continuous hot-upgrade. Without VM startup or shutdown, Vmem’s allocated memory remains static, accessed only by VMs, with no further Vmem operations. The upgrade then involves only interface replacement and atomic waiting, with an average overhead of 2.1\textmu s (99th percentile: 3.5\textmu s). Concurrent VM operations (startup/shutdown) introduce minor delays from mutex locks between memory allocation/release and upgrade tasks. Even then, the average overhead is 2.3\textmu s (99th percentile: 3.6\textmu s). The negligible increase confirms that Vmem’s live upgrade minimally impacts VM operations (Figure~\ref{fig:upgrade_start}), ensuring seamless cloud service continuity and stability.

\captionsetup[figure]{skip=3pt}

\section{Related Work}
HVO~\cite{HVO} reduces metadata by mapping all \texttt{struct page} entries for a huge page to one physical page, but still incurs overhead and complicates management. Folio~\cite{folio} supports arbitrarily sized huge pages yet retains multi-level metadata, offering limited benefit for coarse-grained VM memory. DMEMFS~\cite{dmemfs} removes \texttt{struct page} structures to save memory but relies on static reservation without dynamic repurposing. All three suffer from inflexible granularity, require extensive kernel changes, and lack hot-upgrade support, impacting stability and maintainability.

Researchers have developed huge page management strategies to enhance memory performance and utilization \cite{agarwal2017thermostat,li2019thinking,kwon2016coordinated,panwar2018making,ren2024mtm,lee2023memtis,park2018gcma}. These approaches still rely on struct page-based management and fail to address metadata overhead. Solutions like Illuminator \cite{panwar2018making}, TCMalloc \cite{hunter2021beyond}, FHPM\cite{li2023fhpm}, and HawkEye \cite{panwar2019hawkeye} enhance huge page allocation but may increase system overhead due to complex tracking mechanisms. Other works target memory fragmentation under different workloads \cite{maas2020learning,panwar2016case,novakovic2019mitigating}. In contrast, Vmem naturally avoids 2MB huge-page fragmentation and tail latency from page faults, while supporting mixed-granularity allocation and cross-granularity memory sharing.

To optimize memory address translation overhead, researchers \cite{zhang2024direct,alverti2020enhancing,margaritov2021ptemagnet,stojkovic2022parallel,skarlatos2020elastic} have explored shared memory translation structures, segment mapping, hash lookup, shared memory mapping, and hardware translation. Other studies address translation overhead through shared translation structures, compiler optimizations, hardware enhancements, and permission/load optimizations \cite{dong2016shared,corbet2022sharing,corbet2022_improving_cpu_utilization,huang2021btmmu,schopp2006resizing,alam2017yourself,gorter2022dangzero,tong2021experiences}. These techniques can further help Vmem reduce page translation metadata and performance overhead.

\section{Discussion}
\subsection{Portability}
Vmem’s well-designed allows porting to different OS kernel versions and hardware architectures with minimal code changes, as shown in Table~\ref{tab:portability}. Results indicate minimal modifications, reflecting low porting cost and easy adaptation. Vmem also supports online hot migration from Hugetlb-based systems to Vmem-enabled platforms, requiring only minor adjustments to VM creation logic in libvirt during migration (Table~\ref{tab:adapt}). These features demonstrate Vmem’s strong portability and support widespread deployment.

\begin{table}[htbp]
\caption{Vmem Portability.}
\label{tab:portability}
\scriptsize
\setlength{\tabcolsep}{4pt}
\begin{tabular}{|p{1.3cm}|p{1.3cm}|p{1.3cm}|p{1.5cm}|p{1.5cm}|}
\toprule
\textbf{4.9→4.19} & \textbf{4.9→5.10} & \textbf{x86→ARM} & \textbf{NUMA 2→4} & \textbf{DPU 1.5→2.0} \\
\midrule
80+ LoC (Page Table) & 380+ LoC (Page API) & 50+ LoC (Arch API) & NUMA Parameters Tuning & No Modification \\
\bottomrule
\end{tabular}
\end{table}

\subsection{Functional Extensions}
While Vmem meets the needs of mainstream VMs with DPU-passthrough devices in cloud servers, it is equally applicable to traditional servers without DPUs. It offers extensible features including on-demand allocation, shared memory, memory compaction, AEP support, and filesystem integration. To support cloud elasticity, Vmem can integrate LRU, SWAP, and other mechanisms at huge-page granularity, tailored for cloud-native environments. These features differ from similar OS functions by operating on huge-page-granularity reserved memory and being specifically designed for cloud computing scenarios. Most are already stably deployed in production, with future work aiming to integrate Vmem with CXL for cross-service memory sharing.

\subsection{Hot-Upgrade Experience}
Vmem’s decoupled architecture separates external interfaces from core functions, enabling hot upgrades of kernel modules, and has been deployed in production for upgrading KVM, VFIO, and custom modules. Ensuring data structure compatibility keeps upgrades lightweight and minimally disruptive, and has become the standard for kernel module upgrades in our cloud. For QEMU process hot upgrades \cite{zhang2019fast}, the standard approach preserves the VM’s memory to avoid multi-round copying during migration. However, underlying process information such as \texttt{vma} and \texttt{pid} may change, requiring updates to corresponding metadata in Vmem’s fastmap. These updates are handled seamlessly via \texttt{vm\_ops} callbacks associated with the \texttt{vma}, enabling transparent integration with upper-layer QEMU hot upgrades.

\subsection{Commercial Benefits}
Vmem increases sellable memory by eliminating  the \texttt{struct page} overhead in traditional OS memory management. Although the per-page metadata is small (64\,B per 4\,KB page), the cumulative savings are substantial. On 384\,GB servers, early deployments saved approximately 6\,GB of memory metadata. By minimizing host-reserved memory, Vmem further freed over 10\,GB of sellable memory per server (over 2\%). Deploying Vmem across 300,000 servers yields 3\,PB of additional sellable memory. At current, this reduces server procurement costs by over \$90\,M; at a cloud pricing of \$15/GB/year, it enables \$135\,M in annual incremental revenue. Thus, Vmem not only addresses critical challenges in cloud memory management but also delivers significant commercial value. 

\section{Conclusion}
We present Vmem, a modular, lightweight, and hot-upgradable memory management architecture. By leveraging reserved memory and dynamic kernel module loading, Vmem overcomes limitations of generic OS memory management. Experiments show Vmem significantly improves memory utilization and delivers performance gains in key scenarios. Although implemented in Linux/KVM, its design is applicable to other OSes and virtualization architectures.

\bibliographystyle{ACM-Reference-Format}
\bibliography{vmem}


\begin{thebibliography}{52}


\ifx \showCODEN    \undefined \def \showCODEN     #1{\unskip}     \fi
\ifx \showISBNx    \undefined \def \showISBNx     #1{\unskip}     \fi
\ifx \showISBNxiii \undefined \def \showISBNxiii  #1{\unskip}     \fi
\ifx \showISSN     \undefined \def \showISSN      #1{\unskip}     \fi
\ifx \showLCCN     \undefined \def \showLCCN      #1{\unskip}     \fi
\ifx \shownote     \undefined \def \shownote      #1{#1}          \fi
\ifx \showarticletitle \undefined \def \showarticletitle #1{#1}   \fi
\ifx \showURL      \undefined \def \showURL       {\relax}        \fi
\providecommand\bibfield[2]{#2}
\providecommand\bibinfo[2]{#2}
\providecommand\natexlab[1]{#1}
\providecommand\showeprint[2][]{arXiv:#2}

\bibitem[Agarwal and Wenisch(2017)]%
        {agarwal2017thermostat}
\bibfield{author}{\bibinfo{person}{Neha Agarwal} {and} \bibinfo{person}{Thomas~F Wenisch}.} \bibinfo{year}{2017}\natexlab{}.
\newblock \showarticletitle{Thermostat: Application-transparent page management for two-tiered main memory}. In \bibinfo{booktitle}{\emph{Proceedings of the Twenty-Second International Conference on Architectural Support for Programming Languages and Operating Systems}}. \bibinfo{pages}{631--644}.
\newblock


\bibitem[Alam et~al\mbox{.}(2017)]%
        {alam2017yourself}
\bibfield{author}{\bibinfo{person}{Hanna Alam}, \bibinfo{person}{Tianhao Zhang}, \bibinfo{person}{Mattan Erez}, {and} \bibinfo{person}{Yoav Etsion}.} \bibinfo{year}{2017}\natexlab{}.
\newblock \showarticletitle{Do-it-yourself virtual memory translation}.
\newblock \bibinfo{journal}{\emph{ACM SIGARCH Computer Architecture News}} \bibinfo{volume}{45}, \bibinfo{number}{2} (\bibinfo{year}{2017}), \bibinfo{pages}{457--468}.
\newblock


\bibitem[Alverti et~al\mbox{.}(2020)]%
        {alverti2020enhancing}
\bibfield{author}{\bibinfo{person}{Chloe Alverti}, \bibinfo{person}{Stratos Psomadakis}, \bibinfo{person}{Vasileios Karakostas}, \bibinfo{person}{Jayneel Gandhi}, \bibinfo{person}{Konstantinos Nikas}, \bibinfo{person}{Georgios Goumas}, {and} \bibinfo{person}{Nectarios Koziris}.} \bibinfo{year}{2020}\natexlab{}.
\newblock \showarticletitle{Enhancing and exploiting contiguity for fast memory virtualization}. In \bibinfo{booktitle}{\emph{2020 ACM/IEEE 47th Annual International Symposium on Computer Architecture (ISCA)}}. IEEE, \bibinfo{pages}{515--528}.
\newblock


\bibitem[Amit et~al\mbox{.}(2015)]%
        {amit2015bare}
\bibfield{author}{\bibinfo{person}{Nadav Amit}, \bibinfo{person}{Abel Gordon}, \bibinfo{person}{Nadav Har'El}, \bibinfo{person}{Muli Ben-Yehuda}, \bibinfo{person}{Alex Landau}, \bibinfo{person}{Assaf Schuster}, {and} \bibinfo{person}{Dan Tsafrir}.} \bibinfo{year}{2015}\natexlab{}.
\newblock \showarticletitle{Bare-metal performance for virtual machines with exitless interrupts}.
\newblock \bibinfo{journal}{\emph{Commun. ACM}} \bibinfo{volume}{59}, \bibinfo{number}{1} (\bibinfo{year}{2015}), \bibinfo{pages}{108--116}.
\newblock


\bibitem[Blog(2014)]%
        {kpatch14}
\bibfield{author}{\bibinfo{person}{Red Hat Enterprise~Linux Blog}.} \bibinfo{year}{2014}\natexlab{}.
\newblock \bibinfo{title}{Introducing kpatch: dynamic kernel patching}.
\newblock \bibinfo{numpages}{vol. 26}~pages.
\newblock


\bibitem[Chatzoglou et~al\mbox{.}(2024)]%
        {chatzoglou2024keep}
\bibfield{author}{\bibinfo{person}{Efstratios Chatzoglou}, \bibinfo{person}{Vyron Kampourakis}, \bibinfo{person}{Zisis Tsiatsikas}, \bibinfo{person}{Georgios Karopoulos}, {and} \bibinfo{person}{Georgios Kambourakis}.} \bibinfo{year}{2024}\natexlab{}.
\newblock \showarticletitle{Keep your memory dump shut: Unveiling data leaks in password managers}. In \bibinfo{booktitle}{\emph{IFIP International Conference on ICT Systems Security and Privacy Protection}}. Springer, \bibinfo{pages}{61--75}.
\newblock


\bibitem[Corbet(2022)]%
        {corbet2022sharing}
\bibfield{author}{\bibinfo{person}{Jonathan Corbet}.} \bibinfo{year}{2022}\natexlab{}.
\newblock \bibinfo{title}{Sharing page tables with mshare()}.
\newblock
\urldef\tempurl%
\url{https://lwn.net/Articles/895217/}
\showURL{%
\tempurl}


\bibitem[Corbet(2021)]%
        {folio}
\bibfield{author}{\bibinfo{person}{Jonathan Corbet}.} \bibinfo{year}{2021}\natexlab{}.
\newblock \bibinfo{title}{Clarifying memory management with page folios}.
\newblock
\newblock
\shownote{\url{https://lwn.net/Articles/849538/}}.


\bibitem[Corbet(2022)]%
        {corbet2022_improving_cpu_utilization}
\bibfield{author}{\bibinfo{person}{Jonathan Corbet}.} \bibinfo{year}{2022}\natexlab{}.
\newblock \bibinfo{booktitle}{\emph{Improving {CPU} utilization with the proxy execution scheduler}}.
\newblock LWN.net.
\newblock
\urldef\tempurl%
\url{https://lwn.net/Articles/901059/}
\showURL{%
\tempurl}
\newblock
\shownote{Accessed: 2024-04-01}.


\bibitem[Dong et~al\mbox{.}(2016)]%
        {dong2016shared}
\bibfield{author}{\bibinfo{person}{Xiaowan Dong}, \bibinfo{person}{Sandhya Dwarkadas}, {and} \bibinfo{person}{Alan~L Cox}.} \bibinfo{year}{2016}\natexlab{}.
\newblock \showarticletitle{Shared address translation revisited}. In \bibinfo{booktitle}{\emph{Proceedings of the Eleventh European Conference on Computer Systems}}. \bibinfo{pages}{1--15}.
\newblock


\bibitem[Dong et~al\mbox{.}(2012)]%
        {dong2012high}
\bibfield{author}{\bibinfo{person}{Yaozu Dong}, \bibinfo{person}{Xiaowei Yang}, \bibinfo{person}{Jianhui Li}, \bibinfo{person}{Guangdeng Liao}, \bibinfo{person}{Kun Tian}, {and} \bibinfo{person}{Haibing Guan}.} \bibinfo{year}{2012}\natexlab{}.
\newblock \showarticletitle{High performance network virtualization with SR-IOV}.
\newblock \bibinfo{journal}{\emph{J. Parallel and Distrib. Comput.}} \bibinfo{volume}{72}, \bibinfo{number}{11} (\bibinfo{year}{2012}), \bibinfo{pages}{1471--1480}.
\newblock


\bibitem[Gorman(2004)]%
        {gorman2004understanding}
\bibfield{author}{\bibinfo{person}{Mel Gorman}.} \bibinfo{year}{2004}\natexlab{}.
\newblock \bibinfo{booktitle}{\emph{Understanding the Linux virtual memory manager}}. Vol.~\bibinfo{volume}{352}.
\newblock \bibinfo{publisher}{Prentice Hall Upper Saddle River}.
\newblock


\bibitem[Gorter et~al\mbox{.}(2022)]%
        {gorter2022dangzero}
\bibfield{author}{\bibinfo{person}{Floris Gorter}, \bibinfo{person}{Koen Koning}, \bibinfo{person}{Herbert Bos}, {and} \bibinfo{person}{Cristiano Giuffrida}.} \bibinfo{year}{2022}\natexlab{}.
\newblock \showarticletitle{Dangzero: Efficient use-after-free detection via direct page table access}. In \bibinfo{booktitle}{\emph{Proceedings of the 2022 ACM SIGSAC Conference on Computer and Communications Security}}. \bibinfo{pages}{1307--1322}.
\newblock


\bibitem[Huang et~al\mbox{.}(2016)]%
        {huang2016evolutionary}
\bibfield{author}{\bibinfo{person}{Jian Huang}, \bibinfo{person}{Moinuddin~K Qureshi}, {and} \bibinfo{person}{Karsten Schwan}.} \bibinfo{year}{2016}\natexlab{}.
\newblock \showarticletitle{An evolutionary study of Linux memory management for fun and profit}. In \bibinfo{booktitle}{\emph{2016 USENIX Annual Technical Conference (USENIX ATC 16)}}. \bibinfo{pages}{465--478}.
\newblock


\bibitem[Huang et~al\mbox{.}(2021)]%
        {huang2021btmmu}
\bibfield{author}{\bibinfo{person}{Kele Huang}, \bibinfo{person}{Fuxin Zhang}, \bibinfo{person}{Cun Li}, \bibinfo{person}{Gen Niu}, \bibinfo{person}{Junrong Wu}, {and} \bibinfo{person}{Tianyi Liu}.} \bibinfo{year}{2021}\natexlab{}.
\newblock \showarticletitle{BTMMU: an efficient and versatile cross-ISA memory virtualization}. In \bibinfo{booktitle}{\emph{Proceedings of the 17th ACM SIGPLAN/SIGOPS International Conference on Virtual Execution Environments}}. \bibinfo{pages}{71--83}.
\newblock


\bibitem[Hunter et~al\mbox{.}(2021)]%
        {hunter2021beyond}
\bibfield{author}{\bibinfo{person}{A.H. Hunter}, \bibinfo{person}{Chris Kennelly}, \bibinfo{person}{Paul Turner}, \bibinfo{person}{Darryl Gove}, \bibinfo{person}{Tipp Moseley}, {and} \bibinfo{person}{Parthasarathy Ranganathan}.} \bibinfo{year}{2021}\natexlab{}.
\newblock \showarticletitle{Beyond malloc efficiency to fleet efficiency: a hugepage-aware memory allocator}. In \bibinfo{booktitle}{\emph{15th {USENIX} Symposium on Operating Systems Design and Implementation ({OSDI} 21)}}. \bibinfo{publisher}{{USENIX} Association}, \bibinfo{pages}{257--273}.
\newblock
\showISBNx{978-1-939133-22-9}
\urldef\tempurl%
\url{https://www.usenix.org/conference/osdi21/presentation/hunter}
\showURL{%
\tempurl}


\bibitem[Jain and Choudhary(2016)]%
        {jain2016overview}
\bibfield{author}{\bibinfo{person}{Nancy Jain} {and} \bibinfo{person}{Sakshi Choudhary}.} \bibinfo{year}{2016}\natexlab{}.
\newblock \showarticletitle{Overview of virtualization in cloud computing}. In \bibinfo{booktitle}{\emph{2016 Symposium on Colossal Data Analysis and Networking (CDAN)}}. IEEE, \bibinfo{pages}{1--4}.
\newblock


\bibitem[Jia et~al\mbox{.}(2023)]%
        {jia2023making}
\bibfield{author}{\bibinfo{person}{Weiwei Jia}, \bibinfo{person}{Jiyuan Zhang}, \bibinfo{person}{Jianchen Shan}, {and} \bibinfo{person}{Xiaoning Ding}.} \bibinfo{year}{2023}\natexlab{}.
\newblock \showarticletitle{Making dynamic page coalescing effective on virtualized clouds}. In \bibinfo{booktitle}{\emph{Proceedings of the Eighteenth European Conference on Computer Systems}}. \bibinfo{pages}{298--313}.
\newblock


\bibitem[Jung et~al\mbox{.}(2014)]%
        {jung2014automated}
\bibfield{author}{\bibinfo{person}{Changhee Jung}, \bibinfo{person}{Sangho Lee}, \bibinfo{person}{Easwaran Raman}, {and} \bibinfo{person}{Santosh Pande}.} \bibinfo{year}{2014}\natexlab{}.
\newblock \showarticletitle{Automated memory leak detection for production use}. In \bibinfo{booktitle}{\emph{Proceedings of the 36th International Conference on Software Engineering}}. \bibinfo{pages}{825--836}.
\newblock


\bibitem[Kim et~al\mbox{.}(2015)]%
        {kim2015controlling}
\bibfield{author}{\bibinfo{person}{Sang-Hoon Kim}, \bibinfo{person}{Sejun Kwon}, \bibinfo{person}{Jin-Soo Kim}, {and} \bibinfo{person}{Jinkyu Jeong}.} \bibinfo{year}{2015}\natexlab{}.
\newblock \showarticletitle{Controlling physical memory fragmentation in mobile systems}.
\newblock \bibinfo{journal}{\emph{ACM SIGPLAN Notices}} \bibinfo{volume}{50}, \bibinfo{number}{11} (\bibinfo{year}{2015}), \bibinfo{pages}{1--14}.
\newblock


\bibitem[Kwon et~al\mbox{.}(2023)]%
        {kwon2023efficient}
\bibfield{author}{\bibinfo{person}{Woosuk Kwon}, \bibinfo{person}{Zhuohan Li}, \bibinfo{person}{Siyuan Zhuang}, \bibinfo{person}{Ying Sheng}, \bibinfo{person}{Lianmin Zheng}, \bibinfo{person}{Cody~Hao Yu}, \bibinfo{person}{Joseph Gonzalez}, \bibinfo{person}{Hao Zhang}, {and} \bibinfo{person}{Ion Stoica}.} \bibinfo{year}{2023}\natexlab{}.
\newblock \showarticletitle{Efficient memory management for large language model serving with pagedattention}. In \bibinfo{booktitle}{\emph{Proceedings of the 29th Symposium on Operating Systems Principles}}. \bibinfo{pages}{611--626}.
\newblock


\bibitem[Kwon et~al\mbox{.}(2016)]%
        {kwon2016coordinated}
\bibfield{author}{\bibinfo{person}{Youngjin Kwon}, \bibinfo{person}{Hangchen Yu}, \bibinfo{person}{Simon Peter}, \bibinfo{person}{Christopher~J Rossbach}, {and} \bibinfo{person}{Emmett Witchel}.} \bibinfo{year}{2016}\natexlab{}.
\newblock \showarticletitle{Coordinated and efficient huge page management with ingens}. In \bibinfo{booktitle}{\emph{12th USENIX Symposium on Operating Systems Design and Implementation (OSDI 16)}}. \bibinfo{pages}{705--721}.
\newblock


\bibitem[Lee et~al\mbox{.}(2023)]%
        {lee2023memtis}
\bibfield{author}{\bibinfo{person}{Taehyung Lee}, \bibinfo{person}{Sumit~Kumar Monga}, \bibinfo{person}{Changwoo Min}, {and} \bibinfo{person}{Young~Ik Eom}.} \bibinfo{year}{2023}\natexlab{}.
\newblock \showarticletitle{Memtis: Efficient memory tiering with dynamic page classification and page size determination}. In \bibinfo{booktitle}{\emph{Proceedings of the 29th Symposium on Operating Systems Principles}}. \bibinfo{pages}{17--34}.
\newblock


\bibitem[Li et~al\mbox{.}(2023)]%
        {li2023fhpm}
\bibfield{author}{\bibinfo{person}{Chuandong Li}, \bibinfo{person}{Sai Sha}, \bibinfo{person}{Yangqing Zeng}, \bibinfo{person}{Xiran Yang}, \bibinfo{person}{Yingwei Luo}, \bibinfo{person}{Xiaolin Wang}, {and} \bibinfo{person}{Zhenlin Wang}.} \bibinfo{year}{2023}\natexlab{}.
\newblock \showarticletitle{FHPM: Fine-grained Huge Page Management For Virtualization}.
\newblock \bibinfo{journal}{\emph{arXiv preprint arXiv:2307.10618}} (\bibinfo{year}{2023}).
\newblock


\bibitem[Li(2020)]%
        {boot20}
\bibfield{author}{\bibinfo{person}{Weinan Li}.} \bibinfo{year}{2020}\natexlab{}.
\newblock \showarticletitle{The Practice Method to Speed Up 10x Boot-up Time for Guest in Alibaba Cloud}. In \bibinfo{booktitle}{\emph{KVM Forum 2020}}.
\newblock


\bibitem[Li et~al\mbox{.}(2019)]%
        {li2019thinking}
\bibfield{author}{\bibinfo{person}{Xinyu Li}, \bibinfo{person}{Lei Liu}, \bibinfo{person}{Shengjie Yang}, \bibinfo{person}{Lu Peng}, {and} \bibinfo{person}{Jiefan Qiu}.} \bibinfo{year}{2019}\natexlab{}.
\newblock \showarticletitle{Thinking about a new mechanism for huge page management}. In \bibinfo{booktitle}{\emph{Proceedings of the 10th ACM SIGOPS Asia-Pacific Workshop on Systems}}. \bibinfo{pages}{40--46}.
\newblock


\bibitem[Litke(2007)]%
        {litke2007turning}
\bibfield{author}{\bibinfo{person}{Adam~G Litke}.} \bibinfo{year}{2007}\natexlab{}.
\newblock \showarticletitle{“Turning the Page” on Hugetlb Interfaces}. In \bibinfo{booktitle}{\emph{Proceedings of the Linux Symposium}}. \bibinfo{pages}{277--284}.
\newblock


\bibitem[Lu et~al\mbox{.}(2006)]%
        {lu2006using}
\bibfield{author}{\bibinfo{person}{HJ Lu}, \bibinfo{person}{Kshitij Doshi}, \bibinfo{person}{Rohit Seth}, {and} \bibinfo{person}{Jantz Tran}.} \bibinfo{year}{2006}\natexlab{}.
\newblock \showarticletitle{Using hugetlbfs for mapping application text regions}. In \bibinfo{booktitle}{\emph{Proceedings of the Linux Symposium}}, Vol.~\bibinfo{volume}{2}. \bibinfo{pages}{75--82}.
\newblock


\bibitem[Maas et~al\mbox{.}(2020)]%
        {maas2020learning}
\bibfield{author}{\bibinfo{person}{Martin Maas}, \bibinfo{person}{David~G Andersen}, \bibinfo{person}{Michael Isard}, \bibinfo{person}{Mohammad~Mahdi Javanmard}, \bibinfo{person}{Kathryn~S McKinley}, {and} \bibinfo{person}{Colin Raffel}.} \bibinfo{year}{2020}\natexlab{}.
\newblock \showarticletitle{Learning-based memory allocation for C++ server workloads}. In \bibinfo{booktitle}{\emph{Proceedings of the Twenty-Fifth International Conference on Architectural Support for Programming Languages and Operating Systems}}. \bibinfo{pages}{541--556}.
\newblock


\bibitem[Margaritov et~al\mbox{.}(2021)]%
        {margaritov2021ptemagnet}
\bibfield{author}{\bibinfo{person}{Artemiy Margaritov}, \bibinfo{person}{Dmitrii Ustiugov}, \bibinfo{person}{Amna Shahab}, {and} \bibinfo{person}{Boris Grot}.} \bibinfo{year}{2021}\natexlab{}.
\newblock \showarticletitle{Ptemagnet: Fine-grained physical memory reservation for faster page walks in public clouds}. In \bibinfo{booktitle}{\emph{Proceedings of the 26th ACM International Conference on Architectural Support for Programming Languages and Operating Systems}}. \bibinfo{pages}{211--223}.
\newblock


\bibitem[Mi et~al\mbox{.}(2019)]%
        {mi2019skybridge}
\bibfield{author}{\bibinfo{person}{Zeyu Mi}, \bibinfo{person}{Dingji Li}, \bibinfo{person}{Zihan Yang}, \bibinfo{person}{Xinran Wang}, {and} \bibinfo{person}{Haibo Chen}.} \bibinfo{year}{2019}\natexlab{}.
\newblock \showarticletitle{Skybridge: Fast and secure inter-process communication for microkernels}. In \bibinfo{booktitle}{\emph{Proceedings of the Fourteenth EuroSys Conference 2019}}. \bibinfo{pages}{1--15}.
\newblock


\bibitem[Novakovic et~al\mbox{.}(2019)]%
        {novakovic2019mitigating}
\bibfield{author}{\bibinfo{person}{Stanko Novakovic}, \bibinfo{person}{Alexandros Daglis}, \bibinfo{person}{Dmitrii Ustiugov}, \bibinfo{person}{Edouard Bugnion}, \bibinfo{person}{Babak Falsafi}, {and} \bibinfo{person}{Boris Grot}.} \bibinfo{year}{2019}\natexlab{}.
\newblock \showarticletitle{Mitigating load imbalance in distributed data serving with rack-scale memory pooling}.
\newblock \bibinfo{journal}{\emph{ACM Transactions on Computer Systems (TOCS)}} \bibinfo{volume}{36}, \bibinfo{number}{2} (\bibinfo{year}{2019}), \bibinfo{pages}{1--37}.
\newblock


\bibitem[Panwar et~al\mbox{.}(2019)]%
        {panwar2019hawkeye}
\bibfield{author}{\bibinfo{person}{Ashish Panwar}, \bibinfo{person}{Sorav Bansal}, {and} \bibinfo{person}{K Gopinath}.} \bibinfo{year}{2019}\natexlab{}.
\newblock \showarticletitle{Hawkeye: Efficient fine-grained os support for huge pages}. In \bibinfo{booktitle}{\emph{Proceedings of the Twenty-Fourth International Conference on Architectural Support for Programming Languages and Operating Systems}}. \bibinfo{pages}{347--360}.
\newblock


\bibitem[Panwar et~al\mbox{.}(2016)]%
        {panwar2016case}
\bibfield{author}{\bibinfo{person}{Ashish Panwar}, \bibinfo{person}{Naman Patel}, {and} \bibinfo{person}{K Gopinath}.} \bibinfo{year}{2016}\natexlab{}.
\newblock \showarticletitle{A case for protecting huge pages from the kernel}. In \bibinfo{booktitle}{\emph{Proceedings of the 7th ACM SIGOPS Asia-Pacific Workshop on Systems}}. \bibinfo{pages}{1--8}.
\newblock


\bibitem[Panwar et~al\mbox{.}(2018)]%
        {panwar2018making}
\bibfield{author}{\bibinfo{person}{Ashish Panwar}, \bibinfo{person}{Aravinda Prasad}, {and} \bibinfo{person}{K Gopinath}.} \bibinfo{year}{2018}\natexlab{}.
\newblock \showarticletitle{Making huge pages actually useful}. In \bibinfo{booktitle}{\emph{Proceedings of the Twenty-Third International Conference on Architectural Support for Programming Languages and Operating Systems}}. \bibinfo{pages}{679--692}.
\newblock


\bibitem[Park et~al\mbox{.}(2018)]%
        {park2018gcma}
\bibfield{author}{\bibinfo{person}{SeongJae Park}, \bibinfo{person}{Minchan Kim}, {and} \bibinfo{person}{Heon~Y Yeom}.} \bibinfo{year}{2018}\natexlab{}.
\newblock \showarticletitle{GCMA: Guaranteed contiguous memory allocator}.
\newblock \bibinfo{journal}{\emph{IEEE Trans. Comput.}} \bibinfo{volume}{68}, \bibinfo{number}{3} (\bibinfo{year}{2018}), \bibinfo{pages}{390--401}.
\newblock


\bibitem[Ren et~al\mbox{.}(2024)]%
        {ren2024mtm}
\bibfield{author}{\bibinfo{person}{Jie Ren}, \bibinfo{person}{Dong Xu}, \bibinfo{person}{Junhee Ryu}, \bibinfo{person}{Kwangsik Shin}, \bibinfo{person}{Daewoo Kim}, {and} \bibinfo{person}{Dong Li}.} \bibinfo{year}{2024}\natexlab{}.
\newblock \showarticletitle{MTM: Rethinking memory profiling and migration for multi-tiered large memory}. In \bibinfo{booktitle}{\emph{Proceedings of the Nineteenth European Conference on Computer Systems}}. \bibinfo{pages}{803--817}.
\newblock


\bibitem[Schopp et~al\mbox{.}(2006)]%
        {schopp2006resizing}
\bibfield{author}{\bibinfo{person}{Joel~H Schopp}, \bibinfo{person}{Keir Fraser}, {and} \bibinfo{person}{Martine~J Silbermann}.} \bibinfo{year}{2006}\natexlab{}.
\newblock \showarticletitle{Resizing memory with balloons and hotplug}. In \bibinfo{booktitle}{\emph{Proceedings of the Linux Symposium}}, Vol.~\bibinfo{volume}{2}. \bibinfo{pages}{313--319}.
\newblock


\bibitem[Schrammel et~al\mbox{.}(2022)]%
        {schrammel2022jenny}
\bibfield{author}{\bibinfo{person}{David Schrammel}, \bibinfo{person}{Samuel Weiser}, \bibinfo{person}{Richard Sadek}, {and} \bibinfo{person}{Stefan Mangard}.} \bibinfo{year}{2022}\natexlab{}.
\newblock \showarticletitle{Jenny: Securing Syscalls for {PKU-based} Memory Isolation Systems}. In \bibinfo{booktitle}{\emph{31st USENIX Security Symposium (USENIX Security 22)}}. \bibinfo{pages}{936--952}.
\newblock


\bibitem[Sha et~al\mbox{.}(2024)]%
        {Tiered-Memory-Management}
\bibfield{author}{\bibinfo{person}{Sai Sha}, \bibinfo{person}{Chuandong Li}, \bibinfo{person}{Xiaolin Wang}, \bibinfo{person}{Zhenlin Wang}, {and} \bibinfo{person}{Yingwei Luo}.} \bibinfo{year}{2024}\natexlab{}.
\newblock \showarticletitle{Hardware-Software Collaborative Tiered-Memory Management Framework for Virtualization}.
\newblock \bibinfo{journal}{\emph{ACM Trans. Comput. Syst.}} \bibinfo{volume}{42}, \bibinfo{number}{1–2}, Article \bibinfo{articleno}{4} (\bibinfo{date}{Feb.} \bibinfo{year}{2024}), \bibinfo{numpages}{32}~pages.
\newblock
\showISSN{0734-2071}


\bibitem[Shang et~al\mbox{.}(2023)]%
        {10125028}
\bibfield{author}{\bibinfo{person}{Xiaowei Shang}, \bibinfo{person}{Weiwei Jia}, \bibinfo{person}{Jianchen Shan}, \bibinfo{person}{Xiaoning Ding}, {and} \bibinfo{person}{Cristian Borcea}.} \bibinfo{year}{2023}\natexlab{}.
\newblock \showarticletitle{Reestablishing Page Placement Mechanisms for Nested Virtualization}.
\newblock \bibinfo{journal}{\emph{IEEE Transactions on Cloud Computing}} \bibinfo{volume}{11}, \bibinfo{number}{3} (\bibinfo{year}{2023}), \bibinfo{pages}{3239--3250}.
\newblock


\bibitem[Skarlatos et~al\mbox{.}(2020)]%
        {skarlatos2020elastic}
\bibfield{author}{\bibinfo{person}{Dimitrios Skarlatos}, \bibinfo{person}{Apostolos Kokolis}, \bibinfo{person}{Tianyin Xu}, {and} \bibinfo{person}{Josep Torrellas}.} \bibinfo{year}{2020}\natexlab{}.
\newblock \showarticletitle{Elastic cuckoo page tables: Rethinking virtual memory translation for parallelism}. In \bibinfo{booktitle}{\emph{Proceedings of the Twenty-Fifth International Conference on Architectural Support for Programming Languages and Operating Systems}}. \bibinfo{pages}{1093--1108}.
\newblock


\bibitem[Song(2020)]%
        {HVO}
\bibfield{author}{\bibinfo{person}{Muchun Song}.} \bibinfo{year}{2020}\natexlab{}.
\newblock \bibinfo{title}{Free some vmemmap pages of HugeTLB page}.
\newblock
\newblock
\shownote{\url{https://lwn.net/ml/linux-kernel/20201210035526.38938-1-songmuchun@bytedance.com/}}.


\bibitem[Stojkovic et~al\mbox{.}(2022)]%
        {stojkovic2022parallel}
\bibfield{author}{\bibinfo{person}{Jovan Stojkovic}, \bibinfo{person}{Dimitrios Skarlatos}, \bibinfo{person}{Apostolos Kokolis}, \bibinfo{person}{Tianyin Xu}, {and} \bibinfo{person}{Josep Torrellas}.} \bibinfo{year}{2022}\natexlab{}.
\newblock \showarticletitle{Parallel virtualized memory translation with nested elastic cuckoo page tables}. In \bibinfo{booktitle}{\emph{Proceedings of the 27th ACM International Conference on Architectural Support for Programming Languages and Operating Systems}}. \bibinfo{pages}{84--97}.
\newblock


\bibitem[Tong et~al\mbox{.}(2021)]%
        {tong2021experiences}
\bibfield{author}{\bibinfo{person}{Michael~Hao Tong}, \bibinfo{person}{Robert~L. Grossman}, {and} \bibinfo{person}{Haryadi~S. Gunawi}.} \bibinfo{year}{2021}\natexlab{}.
\newblock \showarticletitle{Experiences in Managing the Performance and Reliability of a {Large-Scale} Genomics Cloud Platform}. In \bibinfo{booktitle}{\emph{2021 USENIX Annual Technical Conference (USENIX ATC 21)}}. \bibinfo{pages}{973--988}.
\newblock


\bibitem[Xie et~al\mbox{.}(2022)]%
        {xie2022cetis}
\bibfield{author}{\bibinfo{person}{Mengyao Xie}, \bibinfo{person}{Chenggang Wu}, \bibinfo{person}{Yinqian Zhang}, \bibinfo{person}{Jiali Xu}, \bibinfo{person}{Yuanming Lai}, \bibinfo{person}{Yan Kang}, \bibinfo{person}{Wei Wang}, {and} \bibinfo{person}{Zhe Wang}.} \bibinfo{year}{2022}\natexlab{}.
\newblock \showarticletitle{CETIS: Retrofitting Intel CET for generic and efficient intra-process memory isolation}. In \bibinfo{booktitle}{\emph{Proceedings of the 2022 ACM SIGSAC Conference on Computer and Communications Security}}. \bibinfo{pages}{2989--3002}.
\newblock


\bibitem[Zhang et~al\mbox{.}(2024b)]%
        {zhang2024direct}
\bibfield{author}{\bibinfo{person}{Jiyuan Zhang}, \bibinfo{person}{Weiwei Jia}, \bibinfo{person}{Siyuan Chai}, \bibinfo{person}{Peizhe Liu}, \bibinfo{person}{Jongyul Kim}, {and} \bibinfo{person}{Tianyin Xu}.} \bibinfo{year}{2024}\natexlab{b}.
\newblock \showarticletitle{Direct Memory Translation for Virtualized Clouds}. In \bibinfo{booktitle}{\emph{Proceedings of the 29th ACM International Conference on Architectural Support for Programming Languages and Operating Systems, Volume 2}}. \bibinfo{pages}{287--304}.
\newblock


\bibitem[Zhang et~al\mbox{.}(2023)]%
        {zhang2023partial}
\bibfield{author}{\bibinfo{person}{Mingxing Zhang}, \bibinfo{person}{Teng Ma}, \bibinfo{person}{Jinqi Hua}, \bibinfo{person}{Zheng Liu}, \bibinfo{person}{Kang Chen}, \bibinfo{person}{Ning Ding}, \bibinfo{person}{Fan Du}, \bibinfo{person}{Jinlei Jiang}, \bibinfo{person}{Tao Ma}, {and} \bibinfo{person}{Yongwei Wu}.} \bibinfo{year}{2023}\natexlab{}.
\newblock \showarticletitle{Partial failure resilient memory management system for ({CXL}-based) distributed shared memory}. In \bibinfo{booktitle}{\emph{Proceedings of the 29th Symposium on Operating Systems Principles}}. \bibinfo{pages}{658--674}.
\newblock


\bibitem[Zhang et~al\mbox{.}(2019)]%
        {zhang2019fast}
\bibfield{author}{\bibinfo{person}{Xiantao Zhang}, \bibinfo{person}{Xiao Zheng}, \bibinfo{person}{Zhi Wang}, \bibinfo{person}{Qi Li}, \bibinfo{person}{Junkang Fu}, \bibinfo{person}{Yang Zhang}, {and} \bibinfo{person}{Yibin Shen}.} \bibinfo{year}{2019}\natexlab{}.
\newblock \showarticletitle{Fast and scalable VMM live upgrade in large cloud infrastructure}. In \bibinfo{booktitle}{\emph{Proceedings of the Twenty-Fourth International Conference on Architectural Support for Programming Languages and Operating Systems}}. \bibinfo{pages}{93--105}.
\newblock


\bibitem[Zhang(2020)]%
        {dmemfs}
\bibfield{author}{\bibinfo{person}{Yulei Zhang}.} \bibinfo{year}{2020}\natexlab{}.
\newblock \bibinfo{title}{Enhance memory utilization with DMEMFS}.
\newblock
\newblock
\shownote{\url{https://lwn.net/Articles/839203/}}.


\bibitem[Zhang et~al\mbox{.}(2024a)]%
        {zhang2024hd}
\bibfield{author}{\bibinfo{person}{Zongpu Zhang}, \bibinfo{person}{Jiangtao Chen}, \bibinfo{person}{Banghao Ying}, \bibinfo{person}{Yahui Cao}, \bibinfo{person}{Lingyu Liu}, \bibinfo{person}{Jian Li}, \bibinfo{person}{Xin Zeng}, \bibinfo{person}{Junyuan Wang}, \bibinfo{person}{Weigang Li}, {and} \bibinfo{person}{Haibing Guan}.} \bibinfo{year}{2024}\natexlab{a}.
\newblock \showarticletitle{Hd-iov: Sw-hw co-designed i/o virtualization with scalability and flexibility for hyper-density cloud}. In \bibinfo{booktitle}{\emph{Proceedings of the Nineteenth European Conference on Computer Systems}}. \bibinfo{pages}{834--850}.
\newblock


\bibitem[Zhou et~al\mbox{.}(2020)]%
        {zhou2020kshot}
\bibfield{author}{\bibinfo{person}{Lei Zhou}, \bibinfo{person}{Fengwei Zhang}, \bibinfo{person}{Jinghui Liao}, \bibinfo{person}{Zhengyu Ning}, \bibinfo{person}{Jidong Xiao}, \bibinfo{person}{Kevin Leach}, \bibinfo{person}{Westley Weimer}, {and} \bibinfo{person}{Guojun Wang}.} \bibinfo{year}{2020}\natexlab{}.
\newblock \showarticletitle{KShot: Live kernel patching with SMM and SGX}. In \bibinfo{booktitle}{\emph{2020 50th Annual IEEE/IFIP International Conference on Dependable Systems and Networks (DSN)}}. IEEE, \bibinfo{pages}{1--13}.
\newblock


\end{thebibliography}
\end{document}